%% file: eccpa.tex
\documentclass[prd,amssymb,amsmath,amsfonts,nofootinbib,reprint,showpacs,longbibliography]{revtex4-1}

\usepackage{graphicx}
\usepackage{lmodern}
\usepackage{amsmath,amssymb}
\usepackage{mathrsfs}
\usepackage{amsfonts}
\usepackage[utf8]{inputenc}
\usepackage{url}
\usepackage[colorlinks]{hyperref}
\makeatletter
\let\eccpa@href\href
\renewcommand{\href}[2]{\begingroup\def\_{\string_}\edef\x{\endgroup\noexpand\eccpa@href{\zap@space#1 \@empty}}\x{#2}}
\makeatother
\usepackage[table]{xcolor}
\usepackage{multirow}
\usepackage[normalem]{ulem}
\usepackage{lipsum}
\normalsize
\usepackage{booktabs}
\usepackage{tabularx}
\usepackage{paralist}

\usepackage{marvosym}
\usepackage{enumerate}
\usepackage{color,soul}
\usepackage{acronym}

\usepackage{bm}
\usepackage{color}
\usepackage{commath}
\allowdisplaybreaks
\usepackage{multirow}

\begin{document}
\title{Post-adiabatic effective-one-body waveforms for coalescing compact binaries}

\input{symbols.tex}
\input{acronyms.tex}
\input{pa_vs_ode_numbers}
\input{v6cmp_numbers}
\input{sweep_numbers}
\input{fd_numbers}

\author{Rossella \surname{Gamba}${}^{1}$}
\affiliation{${}^{1}$ Department of Physics, University of California, Berkeley, CA 94720, USA}

\begin{abstract}
    I apply and extend the formalism of osculating orbits of Refs.~\cite{Damour:2004bz,Lynch:2026ibo}
    to eccentric \ac{eob} models, in order to obtain an efficient yet accurate description of the inspiral-to-plunge
    dynamics of tidal, spinning compact binaries, including precessing ones by twisting the otherwise planar motion.
    The method is recast in terms of equinoctial orbital elements, to circumvent singularities associated with the zero-eccentricity limit.
    While the formalism is general and can be applied to any \ac{eob} model, I focus here on the \dali{} family and demonstrate
    how this technique speeds up waveform generation without significantly compromising the accuracy of the model. Over $10^4$ \acp{bns} and as many
    \acp{bbh} with initial eccentricities up to $0.6$ at $20$ Hz, the median speedup against an optimized version of the model 
    is $\sim \SwSpeedupBnsLow$ for \acp{bns} and $\sim \SwSpeedupBbhLow$ for \acp{bbh}, 
    with mismatches against the full numerical solution always below $\SwMmMaxLowBound$.
    The largest speedups, up to $\sim \SwSpeedupBbhMaxLight$ for individual systems, are obtained for light \acp{bbh} ($M \lesssim 8 M_\odot$) with low
    eccentricities $(e_0 \lesssim 0.1)$. In absolute terms, the median time needed to generate a \ac{bns} waveform is $\sim \SwTimePABns$ s,
    with $\SwFracBnsSubsecond\%$ of the systems below one second, and for long signals the model is up to $\sim \VcmpVsVoneMax$ times
    faster than a previous version. Applied to the calculation of initial conditions in the $e \rightarrow 0$ limit, the inclusion of
    radiation-reaction effects allows the model's input eccentricity to agree with the gauge-invariant waveform eccentricity
    down to initial values of $\sim 7 \times 10^{-8}$. Beyond time-domain applications, the
    same construction also gives frequency-domain \ac{eob} waveforms for eccentric
    compact binaries via a multi-harmonic stationary phase approximation.
\end{abstract}
\date{\today}

\maketitle

\acresetall

\section{Introduction}
Models for \acp{gw} from compact objects have now reached a state of maturity,
being able to handle a variety of orbital configurations across different
regimes~\cite{Pompili:2023tna,vandeMeent:2023ols,Nagar:2024oyk,Gamba:2024cvy,Albanesi:2025txj,Ramos-Buades:2021adz,Khalil:2021txt,Gamboa:2024hli,Gamboa:2024imd,Planas:2025feq,Gamboa:2026jht,Pan:2013rra,Babak:2016tgq,Ossokine:2020kjp,Khalil:2023kep,Ramos-Buades:2023ehm,Estelles:2025zah,Akcay:2020qrj,Gamba:2021ydi,Pratten:2020ceb,Estelles:2021gvs,Hamilton:2021pkf}.
Eccentricity in particular has received considerable attention, as a clean tracer of dynamical
formation~\cite{Samsing:2017xmd,Rodriguez:2017pec,Zevin:2021rtf} for which hints have been claimed in a number of
events~\cite{Romero-Shaw:2020thy,Gayathri:2020coq,Gamba:2021gap,Gupte:2024jfe,Jan:2025fps,Planas:2025jny,Planas:2025plq,Xu:2025ajj,Morras:2025xfu,Malagon:2026uev}.
Nonetheless, many investigations are still required before the complete adoption of eccentric models in
large-scale data analysis: challenges in the definition of a common
eccentricity parameter~\cite{Knee:2022hth,Ramos-Buades:2022lgf,Shaikh:2023ypz,Boschini:2024scu} still need to be overcome,
interplays with other effects, such as spin precession or tides (whose resonances eccentricity can
enhance~\cite{Yang:2019kmf,Takatsy:2026doc}) need to be
understood~\cite{Favata:2013rwa,Cho:2022cdy,Huez:2025npe,CalderonBustillo:2020xms,Romero-Shaw:2022fbf,Xu:2022zza,Divyajyoti:2025cwq,Tibrewal:2026jci,RoyChowdhury:2026xgb},
and the exploration of different parameterizations at the time
of sampling~\cite{Clarke:2022fma,Bonino:2022hkj,Ramos-Buades:2023yhy} might prove beneficial in
\ac{pe} applications~\cite{Lange:2026eqx, Pompili:2026yxq}.

Beyond conceptual issues, the most significant practical roadblock is represented by the
\textit{speed} of the models themselves.
Across families, the introduction of eccentricity comes with a significant hit in waveform generation
efficiency. This is at its core due to the fact that systems of \acp{ode} need to be solved numerically in order to
obtain accurate descriptions of the generic two-body dynamics, with a time-step set by the orbital timescale
rather than by the much longer radiation-reaction one.
Ways to overcome this problem have been developed in the past for quasi-circular binaries. The \ac{pa}
approximation (initially introduced for the computation of eccentricity-reduced initial
data~\cite{Buonanno:2000ef,Damour:2012ky})
was shown to deliver exceptional performance in the description of the inspiral of circularized
systems~\cite{Nagar:2018gnk,Rettegno:2019tzh,Mihaylov:2021bpf,Mihaylov:2023bkc},
directly enabling the usage of \ac{eob} models~\cite{Buonanno:1998gg,Buonanno:2000ef,Damour:2000we,Damour:2001tu} in \ac{gw} data analysis and allowing for
a straightforward integration with likelihood speedup techniques such as relative binning~\cite{Zackay:2018qdy,Leslie:2021ssu,Cornish:2021lje,Huez:2025gja,Wang:2025tnu,Haberland:2026xvj}, multibanding~\cite{Vinciguerra:2017ngf, Morisaki:2021ngj} and reduced-order quadratures~\cite{Canizares:2014fya}.
Alternatives to these analytical techniques include numerical methods such as reduced-order modeling~\cite{Field:2011mf,Purrer:2014fza,Purrer:2015tud,Bohe:2016gbl,Lackey:2016krb,Lackey:2018zvw,Cotesta:2020qhw,Tissino:2022thn}
and surrogate modeling~\cite{Field:2013cfa,Blackman:2017pcm,Varma:2018mmi,Varma:2019csw,Islam:2022laz,Gadre:2022sed,Rink:2024swg,Field:2025isp,Tissino:2022thn,Chua:2018woh,Schmidt:2020yuu,Khan:2020fso,Thomas:2022rmc,GramaxoFreitas:2024bpk,Whittall:2026yvp,Purrer:2026wgp}, the latter of which has recently been
demonstrated to be applicable also to eccentric systems~\cite{Islam:2021mha,Yun:2021jnh,Shi:2024age,Islam:2025llx,Nee:2025nmh,Maurya:2025shc,Islam:2026blk}.

While the \ac{pa} method is not directly applicable to binaries on eccentric orbits, since it requires the
radial separation to decrease monotonically, other techniques
have shown promise in speeding-up the calculation of the radiation-reacted dynamics and waveform.
In 2004, \ac{dgi}~\cite{Damour:2004bz} applied the method of variation of constants to the second-\ac{pn} \ac{eom}
in \ac{adm} coordinates parameterized in terms of quasi-Keplerian elements, obtaining a very
efficient way of generating both dynamics and waveform (see also Refs.~\cite{Memmesheimer:2004cv, Konigsdorffer:2006zt,Cho:2021oai}).
In parallel, the \ac{gsf} community has worked
to tackle the problem of \acp{emri}~\cite{Mino:2003yg,Pound:2007th,Hinderer:2008dm,Barack:2018yvs}, developing
\acp{nit}~\cite{Kevorkian:1987,VanDeMeent:2018cgn,McCart:2021upc,Lynch:2021ogr,Lynch:2023gpu,Drummond:2023wqc} and utilizing interpolation to be able to
obtain waveforms in $\mathcal{O}$(ms)~\cite{Chua:2020stf,Katz:2021yft,Chapman-Bird:2025xtd,Lynch:2026ibo}. Recently, Lynch et al.~\cite{Lynch:2026ibo} have applied
these techniques to an inspiral-only nonspinning \ac{eob} model, obtaining overall speedups up to $\times 8$
with respect to full \ac{ode} solutions, and have since extended them to the circular limit~\cite{Lynch:2026eyw}.
Similarly to the quasi-circular case~\cite{Gamba:2020ljo}, the use of eccentric \ac{pa} techniques
allows for the calculation of waveforms directly in the \ac{fd} via a ``multi-voice''
approach~\cite{Hughes:2021exa,Speri:2023jte,Pound:2021qin}, which reduces to the standard \ac{spa} in the quasi-circular limit.

In this work, I apply the framework of Refs.~\cite{Damour:2004bz,Lynch:2026ibo} to the tidal, spin-precessing \ac{eob} model \dali{},
recasting the method in terms of \textit{equinoctial} orbital elements~\cite{1972CeMec...5..303B,1985CeMec..36..409W,2021CeMDA.133...50B},
for which the circular orbit is a regular point.
This choice allows the method to reach the quasi-circular limit, where the periastron direction, and with it the
anomaly, become ill-defined. 
After demonstrating how the method can be used to compute initial conditions that account for radiation reaction effects at small eccentricities,
I validate and benchmark full evolutions of various binary configurations against their \ac{ode} counterparts.
Then, I illustrate the application of the method to the generation of frequency-domain waveforms.

The paper is organized as follows: in Sec.~\ref{sec:analytical} I summarize the method of
variation of constants (or ``osculating orbits''), following and extending the seminal work of
\ac{dgi}, and introduce the equinoctial elements.
In Sec.~\ref{sec:application} I apply it to \dali{} (Sec.~\ref{sec:dali}), to compute initial conditions (Sec.~\ref{sec:ics})
and evolve the binary systems under radiation reaction effects (Sec.~\ref{sec:benchmarks}), 
highlighting both accuracy and computational efficiency. Then, I compare its performance against other
existing \ac{eob} models (Sec.~\ref{sec:comparison}).
In Sec.~\ref{sec:fd} I sketch the procedure for efficiently generating frequency domain waveforms, bypassing
the need for a full time-domain plus \ac{fft} transformation, discuss delicate aspects of the procedure,
and show its proof-of-concept application to a handful of representative binary configurations.
Finally, in Sec.~\ref{sec:conclusion},
I summarize the findings of this work and outline possible future directions. 
The main paper is complemented by five appendices
which provide additional technical details and clarifications for the main text.

\paragraph*{Notation:} Throughout the paper, I use geometric units $G=c=1$ unless otherwise stated.
I denote the masses of the binary components as $m_1, m_2$, the total mass as $M = m_1 + m_2$, the reduced mass as 
$\mu$, and by $S_1, S_2$ the projections of the spin angular momenta on the orbital angular momentum,
the only spin quantities entering the orbital dynamics (Sec.~\ref{sec:dali}). The mass ratio $q$, the symmetric mass ratio $\nu$ and the dimensionless spins $\chi_{1,2}$ are defined as
\begin{equation}
    q = \frac{m_1}{m_2} \geq 1 , \qquad \nu = \frac{\mu}{M} = \frac{m_1 m_2}{M^2} , \qquad \chi_{1,2} = \frac{S_{1,2}}{m_{1,2}^2} .
\end{equation} The dimensionless tidal polarizability parameters of the two bodies are
$\Lambda_{1,2}$.
While most of the derivations in the paper are valid for any gauge or coordinate choice, the results presented
are tailored to the \ac{eob} formalism. I will in general assume that the orbit is described in \ac{eob}
polar coordinates $(r, \varphi)$, using the momentum conjugate to the tortoise radial coordinate in place of the radial momentum $p_r$,
\begin{equation}
    p_{r_*} = \sqrt{\frac{A}{B}}\, p_r ,
\end{equation}
with $A$ and $B$ the \ac{eob} metric potentials. The arguments of the Hamiltonian are always ordered as $H(r, p_{r_*}, p_\varphi)$,
and subscripts on $H$ denote partial derivatives, so that for instance $H_r$ is its derivative with respect to $r$.
As is customary in \ac{eob} literature, the Hamiltonian and the momenta are rescaled 
to be dimensionless.
Overdots denote derivatives with respect to coordinate time $t$.

\section{Analytical Framework}
\label{sec:analytical}
In this section I briefly summarize and extend the method presented in~\ac{dgi}
by pushing it to higher orders and recasting it in terms of the equinoctial parameterization
of Refs.~\cite{1972CeMec...5..303B,1985CeMec..36..409W,2021CeMDA.133...50B}.
Details of the technical implementation are left to App.~\ref{app:techimp}.

\subsection{Variation of constants and \ac{dgi}}
\label{sec:dgi_summary}

The method that \ac{dgi} builds upon is an extension of the classical method of variation of constants.
This is a standard technique in celestial mechanics, going back to Euler and Lagrange and given its general form
by Lagrange in 1808--1810~\cite{Lagrange:1808,Lagrange:1809,Lagrange:1810,2009LMaPh..90....3M}, and later modified and extended by many authors (see e.g.\ Refs.~\cite{Brouwer:1961,Murray:1999} for textbook
treatments, Refs.~\cite{1972CeMec...5..303B,1985CeMec..36..409W,2021CeMDA.133...50B} for its equinoctial formulation, and
Refs.~\cite{Damour:1983tz,Lincoln:1990ji,Pound:2007th,Gair:2010iv} for its application to relativistic binaries).
It is best known as a way to solve inhomogeneous linear \acp{ode}, but it is in fact an exact change of variables for any
perturbed system: applied to the problem of Keplerian motion, it allows one
to express the solution of the perturbed system in terms of the solution of the unperturbed one,
with the constants of motion replaced by slowly varying functions of time.

\begin{figure}
  \includegraphics[width=\columnwidth]{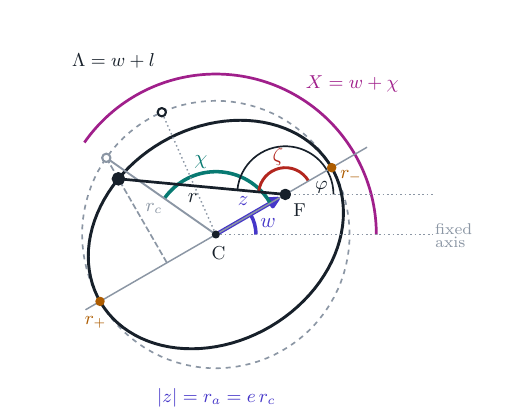}
  \caption{Orbital elements in the Newtonian limit, where each is a length or an angle in the orbital plane. The orbit has its center
  in C and the center of mass of the binary sits at the focus F; $r_c$ is the semi-major axis, $r_a$ the distance between C and F, and
  $r_\pm$ are the turning points; $\chi$ is the eccentric anomaly, $\zeta$ the true anomaly and $l$ the mean anomaly. The
  equinoctial elements of Sec.~\ref{sec:equinoct} are also shown: $z$ is the vector from C to F, $w$ the longitude of periastron
  measured from a fixed axis, $X$ the anomaly measured from the same axis and $\Lambda$ the mean longitude.}
  \label{fig:equinoct_newt}
\end{figure}

Following the notation of \ac{dgi}, consider a two-body system described by a Hamiltonian $H$.
In the conservative case, the \acp{eom} are given by
\begin{equation}
    \label{eq:hamilton}
    \dot{\bm{r}} = \frac{\partial H}{\partial \bm{p}}, \qquad
    \dot{\bm{p}} = -\frac{\partial H}{\partial \bm{r}} ,
\end{equation}
where $\bm{r}$ and $\bm{p}$ are the relative position and momentum of the two bodies, respectively.
Assuming planar motion, it is convenient to work in polar coordinates $(r, \varphi)$, with the momenta
$(p_{r_*}, p_\varphi)$ introduced above.
The system has two constants of motion,
the energy $E$ and the angular momentum $p_\varphi$, which I denote collectively as $c_\mu$.
The bound orbit can then be parameterized in terms of a phase variable $\chi$ that increases monotonically with time, and
\begin{equation}
    \label{eq:chi_param}
    \begin{aligned}
    r &= r_c - r_a \cos \chi, \qquad &e &\equiv \frac{r_a}{r_c} , \\
    r_c &= \frac{r_+ + r_-}{2}, \qquad &r_a &= \frac{r_+ - r_-}{2},
    \end{aligned}
\end{equation}
where $r_\pm$ are the turning points of the motion and $e$ is the eccentricity (see Fig.~\ref{fig:equinoct_newt}). The turning points can
be easily obtained from energy and angular momentum conservation, imposing
\begin{equation}
    H(r_\pm, 0, p_\varphi) = E.
\end{equation}

In this conservative case, the \acp{eom} can be solved via quadratures, leading to a parametric solution
for the motion in terms of $\chi$. The explicit evolution as a function of time
is then reconstructed by integrating $dt/d\chi$ over $\chi$:
\begin{equation}
    \label{eq:dtdchi}
    \frac{dt}{d\chi} = \frac{r_a \sin\chi}{\dot{r}} \, .
\end{equation}
Note that this expression is finite at the turning points, where numerator and denominator vanish together~\cite{Pound:2007th,Hinderer:2017jcs}.
The radial period of the motion is then:
\begin{equation}
    P_r = \int_0^{2\pi} \frac{dt}{d\chi} \, d\chi ,
\end{equation}
while the total change in the azimuthal angle over one radial period is
\begin{equation}
    \Delta \varphi = \int_0^{2\pi} \dot{\varphi}\,\frac{dt}{d\chi} \, d\chi ,
\end{equation}
from which one can define the mean motion $n$ and the periastron advance $k$,
\begin{equation}
    n = \frac{2\pi}{P_r} , \qquad k = \frac{\Delta \varphi}{2\pi} - 1 .
\end{equation}
The conservative motion is quasi-periodic, and it is convenient to introduce two angles that advance uniformly,
\begin{equation}
    \label{eq:l_lambda}
    \dot{l} = n(c_\mu), \qquad \dot{\lambda} = (1+k(c_\mu))\, n(c_\mu) ,
\end{equation}
with $l = 0$ at periastron, in terms of which the radial separation and the azimuthal angle read
\begin{equation}
    \label{eq:SW}
    r = S(l; c_\mu), \qquad \varphi = \lambda + V(l; c_\mu) ,
\end{equation}
with $S$ and $V$ periodic functions of $l$ of period $2\pi$. The angle $l$ follows from $\chi$ by quadrature,
\begin{equation}
    \label{eq:l_of_chi}
    \begin{aligned}
    l(\chi; c_\mu) &= n(c_\mu) \int_0^{\chi} \frac{dt}{d\chi'}\, d\chi' \equiv \chi + T(\chi; c_\mu) , \\
    \frac{dl}{d\chi} &= n\, \frac{dt}{d\chi} ,
    \end{aligned}
\end{equation}
with $T$ periodic in $\chi$ and vanishing at $\chi = 0$, so that
\begin{equation}
    \label{eq:S_of_l}
    \begin{aligned}
    S(l) &= r_c - r_a \cos\chi(l) , \\
    V(l) &= \varphi(\chi(l)) - (1+k)\, l , \qquad
    \varphi(\chi) = \int_0^{\chi} \dot\varphi\, \frac{dt}{d\chi'}\, d\chi' .
    \end{aligned}
\end{equation}
Inverting $l(\chi)$ is the analogue of Kepler's equation. For a generic $H$ it has no closed form, and
$\chi(l)$ is obtained numerically. Averages over an orbit are taken over $l$ or $\chi$ as:
\begin{equation}
    \label{eq:avg_chi}
    \langle g \rangle \equiv \frac{1}{2\pi} \int_0^{2\pi} g\, dl = \frac{1}{P_r} \int_0^{2\pi} g\, \frac{dt}{d\chi}\, d\chi \, .
\end{equation}

When a perturbation is introduced, the constants of motion become functions of time.
Expressing dissipative forces in terms of the radial and angular components $(\mathcal{F}_r, \mathcal{F}_\varphi)$,
the equations of motion read
\begin{equation}
    \label{eq:eom_tortoise}
    \begin{aligned}
    \dot r &= \sqrt{\frac{A}{B}}\, \frac{\partial H}{\partial p_{r_*}} , \qquad
    &\dot p_{r_*} &= \sqrt{\frac{A}{B}} \Bigl( -\frac{\partial H}{\partial r} + \mathcal{F}_r \Bigr) , \\
    \dot\varphi &= \frac{\partial H}{\partial p_\varphi} \equiv \Omega , \qquad
    &\dot p_\varphi &= \mathcal{F}_\varphi .
    \end{aligned}
\end{equation}
The evolution of the now slowly varying constants $c_\mu$ is
trivially obtained from energy balance and the Hamilton equations above:
\begin{equation}
    \label{eq:cdot}
    \dot{E} \equiv F_E = \dot{r} \mathcal{F}_r + \dot{\varphi} \mathcal{F}_\varphi, \qquad
    \dot{p}_\varphi \equiv F_{p_\varphi} = \mathcal{F}_\varphi .
\end{equation}
The angle variables $l$ and $\lambda$, too, drift from their unperturbed values:
\begin{equation}
    \label{eq:ldot_pert}
    \dot{l} = n(c_\mu) + F_l(l; c_\mu), \qquad
    \dot{\lambda} = (1+k(c_\mu)) n(c_\mu) + F_\lambda(l; c_\mu).
\end{equation}
The exact form of $F_l$ and $F_\lambda$ follows from differentiating Eq.~\eqref{eq:SW} with respect to time and
imposing the \textit{osculating orbits} condition, i.e. that the perturbed trajectory be tangent to an
unperturbed one at each instant (App.~\ref{app:osc}). 
With this requirement, one obtains:
\begin{equation}
    \label{eq:Fl_general}
    F_l = -\frac{\partial S}{\partial c_\mu} \dot{c}_\mu \left(\frac{\partial S}{\partial l}\right)^{-1}, \qquad
    F_\lambda = -\frac{\partial V}{\partial l} F_l - \frac{\partial V}{\partial c_\mu} \dot{c}_\mu .
\end{equation}

The reformulation of the \ac{eom} in terms of
the osculating elements is exact, and provides a convenient way
of separating the fast orbital motion from the slow evolution of the orbital elements.
Rewriting the equations with $l$ as the independent variable, one obtains
\begin{equation}
    \label{eq:exact_in_l}
    \begin{aligned}
    \frac{dc_\mu}{dl} &= \frac{F_\mu(l; c_\sigma)}{n(c_\sigma) + F_l(l; c_\sigma)} \, , \\
    \frac{d\lambda}{dl} &= \frac{(1+k) n + F_\lambda(l; c_\sigma)}{n(c_\sigma) + F_l(l; c_\sigma)} \, .
    \end{aligned}
\end{equation}
This system can be further reformulated in terms of $\chi$; derivatives with respect to $l$
transforming according to the second of Eqs.~\eqref{eq:l_of_chi}.

The perturbation provided by the dissipative forces is, by definition, small: for
\ac{gw} emission, the \ac{lo} effect is of order $c^{-5}$. Therefore, it is convenient
to introduce a \ac{pa} bookkeeping parameter $\epsilon \ll 1$, and attach it
to the dissipative forces
\begin{equation}
    F_\mu \to \epsilon F_\mu \, , \qquad F_l \to \epsilon F_l \, , \qquad F_\lambda \to \epsilon F_\lambda \, .
\end{equation}
Expanded in $\epsilon$, the exact equations \eqref{eq:exact_in_l} become:
\begin{equation}
    \begin{aligned}
    \frac{dc_\mu}{dl} &= \frac{\epsilon F_\mu}{n}
                         - \frac{\epsilon^2 F_\mu F_l}{n^2} + O(\epsilon^3) , \\
    \frac{d\lambda}{dl} &= 1 + k + \frac{\epsilon\, [F_\lambda - (1+k) F_l]}{n} + O(\epsilon^2) ,
    \end{aligned}
\end{equation}
and keeping only the first term gives
\begin{equation}
    \label{eq:pa1}
    \begin{aligned}
    \frac{dc_\mu}{dl} &\simeq \frac{ \epsilon F_\mu(l; c_\sigma)}{n(c_\sigma)} \equiv G_\mu(l; c_\sigma), \\
    \frac{d\lambda}{dl} &\simeq 1 + k(c_\sigma) \, .
    \end{aligned}
\end{equation}
The perturbations themselves can be decomposed into a slow, secular part and a fast, oscillating part,
\begin{equation}
    \begin{aligned}
    \bar{G}_\mu(c_\sigma) &= \langle G_\mu \rangle \equiv \frac{1}{2\pi} \int_0^{2\pi} G_\mu(l; c_\sigma)\, dl, \\
    \tilde{G}_\mu(l; c_\sigma) &= G_\mu - \bar{G}_\mu ,
    \end{aligned}
\end{equation}
which induces the corresponding decomposition of the constants of motion,
\begin{equation}
    \label{eq:split}
    c_\mu(l) = \bar{c}_\mu(l) + \epsilon \tilde{c}_\mu(l) \, .
\end{equation}
$\epsilon \tilde{c}_\mu$ represents the fast, oscillating part of the constant of motion,
and is $O(\epsilon)$ relative to the secular part $\bar{c}_\mu$\footnote{This is a consequence of $\tilde{c}_\mu$
being the zero-average primitive of $\tilde{G}_\mu(l; \bar{c}_\sigma)$.}.
I call \textit{mean orbit} the conservative orbit labelled by the $\bar c_\mu$, i.e.\ the one the binary would follow if its
elements were frozen at their secular values; the osculating orbit oscillates about it on the orbital timescale.
In the language of \ac{gsf}, the change of variables from $c_\mu$ to $\bar c_\mu$ is a \ac{nit}: it differs from the identity
by small terms periodic in $l$ (here with zero average), chosen so that, to the order kept, the evolution equations of the new
variables no longer depend on $l$~\cite{VanDeMeent:2018cgn,Lynch:2021ogr,Lynch:2026ibo}\footnote{Eq.~\eqref{eq:split}, which gives the osculating elements in
terms of the mean ones, is the inverse of the usual \ac{nit} as written in the \ac{gsf} literature.}.
Given this expansion, the arguments of $G_\mu(l; c_\sigma)$ and of $k(c_\sigma)$ can be effectively replaced by their
secular parts $\bar{c}_\sigma$ when considering the long-term evolution, up to corrections of order $O(\epsilon)$.

Practically, the system then reads
\begin{equation}
    \label{eq:pa_system}
    \frac{d\bar c_\mu}{dl} = \bar G_\mu(\bar c_\sigma), \quad
    \frac{d\tilde c_\mu}{dl} = \tilde G_\mu(l; \bar c_\sigma), \quad
    \frac{d\lambda}{dl} = 1 + k(\bar c_\sigma) ,
\end{equation}
and the right-hand sides of all three depend only on the slow constants $\bar c_\mu$ (which are obtained by
solving the first equation) or the angle $l$, and \textit{not} the fast oscillating corrections $\tilde c_\mu$.

\subsection{Higher orders}
\label{sec:higherpa}
\Ac{dgi} stops the expansion at first order in $\epsilon$ for the $\bar{c}_\mu$. Clearly, however, it is
possible (if not desirable) to push the expansion to higher orders.

Before proceeding, it is useful to clarify the notation for the different orders in the \ac{pa} expansion.
The \ac{eob} literature typically denotes orders by counting powers of the small parameter $\epsilon$, with ``0\ac{pa}'' corresponding to $\epsilon^0$
(adiabatic order), ``1\ac{pa}'' to $\epsilon^1$, and so on.
Here, I will simply refer to the orders as \ac{lo}, \ac{nlo}, \ac{nnlo} and so on.
These are the 0\ac{pa}, 1\ac{pa} and 2\ac{pa} orders of Ref.~\cite{Lynch:2026ibo}.
The N$^i$\ac{lo} level retains terms up to $\epsilon^{i+1}$ in the
evolution of the $\bar c_\mu$, and up to $\epsilon^{i}$ both in the evolution of the angles and in the map from mean to
osculating elements (see below). Labeling each level by these three orders, (map, rate, phase), \ac{lo}, \ac{nlo} and
\ac{nnlo} are $(0, 1, 0)$, $(1, 2, 1)$ and $(2, 3, 2)$, while the \ac{dgi} system of Eq.~\eqref{eq:pa_system} is $(1, 1, 0)$: the
\ac{lo} evolution of its first and last equations, together with the first-order map of the second.

The procedure to compute any desired order can be schematically summarized as follows:
\begin{enumerate}
    \item Expand the right-hand side of~\eqref{eq:exact_in_l} to the desired order of $\epsilon$;
    \item Perform a \ac{nit} to the desired order, and decompose $c_\mu$ into a slow, secular piece and fast oscillating corrections
    that only depend on the secular part and the anomaly:
    \begin{equation}
        \label{eq:c_of_cbar}
        \begin{aligned}
        c_\mu &= \bar{c}_\mu + \sum_{i \geq 1} \epsilon^i\, \tilde{c}^{(i)}_\mu(l; \bar{c}_\sigma) , \qquad
                 \bigl\langle \tilde{c}^{(i)}_\mu \bigr\rangle = 0 , \\
        \frac{d\bar{c}_\mu}{dl} &= \sum_{i \geq 1} \epsilon^i\, \bar{G}^{(i)}_\mu(\bar{c}_\sigma) \, .
        \end{aligned}
    \end{equation}
    Here $\bar{G}^{(1)}_\mu$ is the $\bar G_\mu$ of Sec.~\ref{sec:dgi_summary}, $\bar{c}_\mu$ is defined to
    be such that $d\bar{c}_\mu/dl$ is independent of $l$, and the split is fixed by requiring that the oscillating
    parts have zero average at every $\bar c_\sigma$.
    Equation~\eqref{eq:c_of_cbar} gives the instantaneous osculating $c_\mu$ from the mean elements $\bar{c}_\mu$.
    The inverse transformation can be determined perturbatively requiring that:
    \begin{equation}
        \label{eq:c_of_cbar_inverse}
        c_\mu\bigl(l; \bar{c}_\sigma(l; c_\rho)\bigr) = c_\mu \, .
    \end{equation}
    Practically, writing
    \begin{equation}
    \label{eq:cbar_of_c}
    \bar{c}_\mu = c_\mu + \sum_{i \geq 1} \epsilon^i\, b^{(i)}_\mu(l; c_\sigma) ,
    \end{equation}
    the condition above fixes the coefficients $b^{(i)}_\mu$: inserting Eq.~\eqref{eq:cbar_of_c} into
    Eq.~\eqref{eq:c_of_cbar} and expanding in $\epsilon$ allows one to solve for $b^{(i)}_\mu$ order by order.
    \item Apply the same transformation to the angle $\lambda$ and to the time $t$:
    \begin{equation}
        \label{eq:nit_angles}
        \begin{aligned}
        \lambda &= \bar\lambda + \sum_{i \geq 1} \epsilon^i\, \tilde\lambda^{(i)}(l; \bar c_\sigma) , \qquad
                  \bigl\langle \tilde\lambda^{(i)} \bigr\rangle = 0 , \\
        t &= \bar t + \sum_{i \geq 1} \epsilon^i\, \tilde t^{(i)}(l; \bar c_\sigma) , \qquad
                  \bigl\langle \tilde t^{(i)} \bigr\rangle = 0 .
        \end{aligned}
    \end{equation}
    Like the $\bar c_\mu$, $\bar\lambda$ and $\bar t$ are defined by the requirement that their rates be independent of $l$.
    Unlike the $\bar c_\mu$ they are not slowly evolving, but they are free of
    oscillations on the orbital timescale.
    \item Plug Eq.~\eqref{eq:c_of_cbar} into the expanded Eq.~\eqref{eq:exact_in_l} (in both the left- and right-hand sides)
    and take orbital averages, to determine the governing equation for $\bar{c}_\mu$ at the various $\epsilon$ orders.
    \item Repeat the previous step with Eq.~\eqref{eq:nit_angles}, using the second of Eqs.~\eqref{eq:exact_in_l} for
    $\lambda$ and the inverse of the first of Eqs.~\eqref{eq:ldot_pert} for $t$, to obtain the equations for
    $d\bar{\lambda}/dl$ and $d\bar{t}/dl$.
\end{enumerate}

It is instructive to carry out the expansion explicitly, step by step, to \ac{nlo}.
Following the list above:
\begin{equation}
    \label{eq:second_order}
    \begin{aligned}
    \frac{dc_\mu}{dl} &= \epsilon\, G_\mu(l; c_\sigma) \bigl[ 1 - \epsilon\, G_l(l; c_\sigma) \bigr] + O(\epsilon^3) , \\
    G_l &\equiv \frac{F_l}{n} , \qquad G_\lambda \equiv \frac{F_\lambda}{n} \, ,
    \end{aligned}
\end{equation}
where, as for $G_\mu$, the forcing terms are measured per unit $l$ rather than per unit time.
The relation between the osculating elements $c_\mu$ and the averaged elements $\bar{c}_\mu$, needed for
instance to recast the initial conditions in terms of the averaged elements, is fixed by the following equations:
\begin{equation}
    \begin{aligned}
    b^{(1)}_\mu &= -\tilde{c}^{(1)}_\mu , \\
    b^{(2)}_\mu &= -\tilde{c}^{(2)}_\mu + \frac{\partial \tilde{c}^{(1)}_\mu}{\partial c_\nu}\, \tilde{c}^{(1)}_\nu \, .
    \end{aligned}
\end{equation}
The functions $G_\mu(l; c_\sigma)$ can be further expanded for small $\epsilon$ following Eq.~\eqref{eq:c_of_cbar}:
\begin{equation}
    \begin{aligned}
    G_\mu(l; c_\sigma) &= G_\mu\bigl(l; \bar{c}_\sigma + \epsilon\, \tilde{c}^{(1)}_\sigma + O(\epsilon^2)\bigr) \\
            &= G_\mu(l; \bar{c}_\sigma) + \epsilon\, \frac{\partial G_\mu}{\partial c_\nu}\, \tilde{c}^{(1)}_\nu + O(\epsilon^2) \, .
    \end{aligned}
\end{equation}
Inserting back into \eqref{eq:second_order}, one obtains:
\begin{equation}
    \begin{aligned}
    &\epsilon \bigl( \bar{G}^{(1)}_\mu + \partial_l \tilde{c}^{(1)}_\mu \bigr)
     + \epsilon^2 \Bigl( \bar{G}^{(2)}_\mu + \partial_l \tilde{c}^{(2)}_\mu
     + \frac{\partial \tilde{c}^{(1)}_\mu}{\partial \bar{c}_\nu}\, \bar{G}^{(1)}_\nu \Bigr) \\
    &\qquad = \epsilon\, G_\mu + \epsilon^2 \Bigl( \frac{\partial G_\mu}{\partial c_\nu}\, \tilde{c}^{(1)}_\nu - G_\mu G_l \Bigr) + O(\epsilon^3) ,
    \end{aligned}
\end{equation}
where it is intended that the r.h.s. is evaluated at $\bar{c}_\sigma$.
On orbit-averaging, two terms drop out,
\begin{equation}
    \bigl\langle \partial_l \tilde c^{(i)}_\mu \bigr\rangle = 0 , \qquad
    \Bigl\langle \frac{\partial \tilde c^{(1)}_\mu}{\partial \bar c_\nu} \Bigr\rangle
    = \frac{\partial \langle \tilde c^{(1)}_\mu \rangle}{\partial \bar c_\nu} = 0 ,
\end{equation}
the first by periodicity of the $\tilde{c}^{(i)}_\mu$, and the second because $\tilde c^{(1)}_\mu$ has zero average by definition. 
At first order in $\epsilon$ one recovers the previously obtained terms, and at second order:
\begin{equation}
\bar{G}^{(2)}_\mu = \Bigl\langle \frac{\partial G_\mu}{\partial c_\nu}\, \tilde{c}^{(1)}_\nu - G_\mu G_l \Bigr\rangle \, .
\end{equation}
The oscillating part of the $\epsilon^2$ terms fixes $\tilde c^{(2)}_\mu$.
The equation for the evolution of $\bar{\lambda}$ is obtained in the same way, expanding the periastron
advance $k$ about the mean elements in the second of Eqs.~\eqref{eq:exact_in_l}. One gets:
\begin{equation}
    \label{eq:lambdabar}
    \frac{d\bar{\lambda}}{dl} = 1 + k(\bar{c}_\sigma)
        + \epsilon \bigl[ \bar G_\lambda - (1+k)\, \bar G_l \bigr] + O(\epsilon^2) ,
\end{equation}
the term linear in $\tilde c^{(1)}_\nu$ that comes from the expansion of $k$ averaging to zero as above.
Finally, once more following the same arguments, for the time variable the evolution equation reads:
\begin{equation}
    \label{eq:tbar}
    \frac{d\bar{t}}{dl} = \frac{1}{n(\bar c_\sigma)} \bigl[ 1 - \epsilon\, \bar G_l \bigr] + O(\epsilon^2) \, .
\end{equation}

\subsection{The $e\rightarrow 0$ limit and equinoctial variables}
\label{sec:equinoct}
The parameterization in terms of $(E, p_\varphi)$ is unfortunately rather ill-suited to representing
systems for which $e\rightarrow 0$. 
This is due to two main reasons: 
first, from a practical standpoint, as $e\rightarrow 0$,
$r_a$ becomes increasingly difficult to determine accurately from energy conservation,
leading to numerical noise in the evaluation of the orbital elements;
second, from a theoretical one, in the low eccentricity limit the anomaly parameter $\chi$
becomes ill-defined, as it depends on the periastron direction, which is ambiguous when the orbit is nearly circular.
The consequence of the second issue is that the evolution equations for the angles become singular in this limit
at higher \ac{pa} orders ($F_l \sim 1/e$), as already noted in Ref.~\cite{Lynch:2026ibo}.
The solution to this issue is well-known in the field of celestial mechanics, and consists of
re-parameterizing the orbit in terms of \textit{equinoctial}
orbital elements~\cite{1972CeMec...5..303B,1985CeMec..36..409W,2021CeMDA.133...50B,2014CeMDA.118..197R} (illustrated
schematically in Fig.~\ref{fig:equinoct_newt}), and directly evolving the Laplace-Runge-Lenz vector in place of
two orbital elements, typically the eccentricity and the periastron phase. These elements are at the core of semi-analytical
satellite theories based on averaging~\cite{Danielson:1995}, and analogous variables, $e \sin w$ and $e \cos w$, were introduced
in Ref.~\cite{Pound:2007th} for the osculating orbits of Schwarzschild spacetime, precisely to deal with small eccentricities\footnote{The same
solution is also proposed in the context of \ac{eob} in footnote 1 of Ref.~\cite{Lynch:2026ibo} and implemented in~\cite{Lynch:2026eyw}; preliminary results were presented in~\cite{LynchCapra29}.}

Following this rationale in the context of the \ac{dgi} framework, I define the complex quantity
\begin{equation}
    z = \alpha + i\beta \equiv r_a\, e^{i w} \, ,
\end{equation}
along with the anomaly parameter $X = w + \chi$.
The physical meaning of these newly introduced variables becomes clear once
the radial coordinate $r$ is expressed as a function of $X$ and $z$:
\begin{equation}
    \label{eq:eq_param}
    r(X) = r_c - \text{Re}\big[z^*\, e^{iX}\big] = r_c - (\alpha \cos X + \beta \sin X) \, ,
\end{equation}
where $z^*$ denotes the complex conjugate of $z$. 
Equation~\eqref{eq:eq_param} shows that $z$ is the complex amplitude of the radial oscillation,
while its argument $w$ determines the position of periastron along $X$ ($r = r_-$ at $X = w$).
In the Newtonian limit, $w$ corresponds to the longitude of periastron measured from a reference direction,
and $z$ is parallel to the Laplace-Runge-Lenz vector (see Fig.~\ref{fig:equinoct_newt}).

Because $X$, $\alpha$, and $\beta$ do not rely on an explicitly defined periastron direction,
all three quantities remain regular in the quasi-circular limit $r_a \to 0$. It is therefore convenient to swap the
slow variables $(E, p_\varphi)$ and angle $\chi$ for $(\alpha, \beta, p_\varphi)$ and $X$.
Under this parameterization, the binary orbit can be evolved following the same scheme
of Sec.~\ref{sec:dgi_summary}, 
provided that one is able to express the osculating orbital elements in terms of $(\alpha, \beta, p_\varphi)$ and $X$,
and that the time evolution of $(\alpha, \beta, p_\varphi)$ is known.

Regarding the first point, given $(\alpha, \beta, p_\varphi)$ one can as before estimate $r_c$
by imposing energy conservation at the turning points, after which $p_{r_*}$ also follows.
It is convenient to define the latter as a product of the derivative of $r$ with respect to
$X$ and a function $W_*$:
\begin{equation}
    \label{eq:prstar_of_X}
    p_{r_*}(X) = \frac{dr}{dX}\, W_*(r) , \qquad \frac{dr}{dX} = \mathrm{Im}\bigl[z^*\, e^{iX}\bigr] \, .
\end{equation}
The first piece, $dr/dX$, contains the zeros of $p_{r_*}$ at the turning points, while $W_*$ does not vanish
as $r_a \to 0$. In this way, the cancellations between vanishing quantities are carried out analytically
rather than numerically, e.g. in the osculating conditions below.

As for the second point, the evolution of the equinoctial parameters can be computed by imposing the osculating orbits
condition from Eqs.~\eqref{eq:eq_param} and~\eqref{eq:prstar_of_X},
requiring that $r$ and $p_{r_*}$ be unchanged by the variation of the elements at fixed $X$ (see App.~\ref{app:osc}).
This gives the following system of equations:
\begin{equation}
    \label{eq:eq_osc}
    \begin{pmatrix} a_{11} & a_{12} \\ a_{21} & a_{22} \end{pmatrix}
    \begin{pmatrix} \dot\alpha \\ \dot\beta \end{pmatrix}
    = \begin{pmatrix} R_1 \\ R_2 \end{pmatrix} ,
    \qquad \dot p_\varphi = \mathcal{F}_\varphi \, ,
\end{equation}
with
\begin{equation}
    \label{eq:eq_osc_coeffs}
    \begin{aligned}
    &a_{11} = \cos X - c_{r_a} \alpha , \quad a_{12} = \sin X - c_{r_a} \beta , \\
    &a_{21} = \sin X + \eta\, \alpha , \quad a_{22} = \eta\, \beta - \cos X , \\
    &R_1 = \frac{\partial r_c}{\partial p_\varphi}\, \mathcal{F}_\varphi , \\
    &R_2 = \frac{1}{W_*} \Bigl[ \sqrt{A/B}\, \mathcal{F}_r
           - \frac{dr}{dX} \frac{\partial W_*}{\partial p_\varphi} \mathcal{F}_\varphi \Bigr] ,
    \end{aligned}
\end{equation}
and
\begin{equation}
    c_{r_a} = \frac{1}{r_a} \frac{\partial r_c}{\partial r_a} , \qquad
    \eta = \frac{dr}{dX}\, \frac{1}{r_a W_*} \frac{\partial W_*}{\partial r_a} ,
\end{equation}
the derivatives of $W_*$ taken at fixed $r$. This system can be inverted to obtain $\dot\alpha$ and $\dot\beta$.

Summarizing, the slow variables $c_\mu$ are $(\alpha, \beta, p_\varphi)$, and the fast angles are
$\lambda$ and the mean longitude $\Lambda \equiv l + w$, which replaces $l$.
The two angles are tied to the anomaly and to the orbital phase by the two periodic maps of the
conservative orbit,
\begin{equation}
    \label{eq:TV_maps}
    \Lambda \equiv l + w = X + T(X; c_\mu) , \qquad \varphi = \lambda + V(X; c_\mu) ,
\end{equation}
where $T$ and $V$ are the functions of Eqs.~\eqref{eq:l_of_chi} and~\eqref{eq:SW}, written in terms of $X$ through
$\chi = X - w$, e.g.\ $T(X; c_\mu) \equiv T(\chi = X - w; c_\mu)$: through $w$, both also depend on the direction of $z$.

Given the rates, the expansion in $\epsilon$ proceeds as in Sec.~\ref{sec:higherpa}, with two differences. First, the fast
angle is $\Lambda$ rather than $l$. Second, the variable left untransformed is the time rather than the anomaly~\cite{Pound:2021qin}, so that the
oscillating corrections of Eq.~\eqref{eq:nit_angles} are carried by the angles, $\tilde\Lambda^{(i)}$ and $\tilde\lambda^{(i)}$
(App.~\ref{app:higher}). 
Leaving the radial phase untransformed instead, as in
Refs.~\cite{VanDeMeent:2018cgn,Lynch:2026ibo}, is equally valid, and in that case Sec.~\ref{sec:higherpa}
applies verbatim with $l \to \Lambda$.
The averages are then over $\bar\Lambda$ at
fixed mean elements, or equivalently time averages over a radial period of the mean orbit.
The resulting averaged system does not depend on $\bar\Lambda$, and since $\bar p_\varphi$ is monotonic it can be used as the
integration variable\footnote{The averaged equations could alternatively be integrated in $t$ or $\bar\Lambda$. $\bar p_\varphi$ is
preferred because it is bounded, removes one equation from the system, and connects more directly to the quasi-circular \ac{pa}
approximation (App.~\ref{app:circular}).}.
It then reads
\begin{equation}
    \label{eq:eq_system}
    \begin{aligned}
    \frac{d\bar z}{d\bar p_\varphi} &= \frac{\dot{\bar z}}{\dot{\bar p}_\varphi} , \\
    \frac{dt}{d\bar p_\varphi} &= \frac{1}{\dot{\bar p}_\varphi} , \qquad
    \frac{d\bar\Lambda}{d\bar p_\varphi} = \frac{n + \bar F_\Lambda}{\dot{\bar p}_\varphi} , \\
    \frac{d\bar\lambda}{d\bar p_\varphi} &= \frac{(1+k)\, n + \bar F_\lambda}{\dot{\bar p}_\varphi} \, ,
    \end{aligned}
\end{equation}
where all quantities on the right-hand side are evaluated on the mean orbit, $\dot{\bar z} = \dot{\bar\alpha} + i\,\dot{\bar\beta}$,
and the rates are expanded as in Eq.~\eqref{eq:c_of_cbar},
\begin{equation}
    \label{eq:eq_rates}
    \begin{aligned}
    \dot{\bar c}_\mu &= \sum_{i \geq 1} \epsilon^i\, \bar F^{(i)}_\mu , \\
    \bar F_\Lambda &= \sum_{i \geq 1} \epsilon^i\, \bar F^{(i)}_\Lambda , \qquad
    \bar F_\lambda = \sum_{i \geq 1} \epsilon^i\, \bar F^{(i)}_\lambda .
    \end{aligned}
\end{equation}
At first order, $\bar F^{(1)}_\mu = \langle F_\mu \rangle$ are the orbit averages of the rates of Eq.~\eqref{eq:eq_osc}, and
\begin{equation}
    \bar F^{(1)}_\Lambda = \Bigl\langle \frac{\partial T}{\partial c_\mu} F_\mu \Bigr\rangle , \qquad
    \bar F^{(1)}_\lambda = -\Bigl\langle \frac{\partial V}{\partial c_\mu} F_\mu \Bigr\rangle
\end{equation}
are 
the analogues of the $F_l$ and $F_\lambda$ of Sec.~\ref{sec:dgi_summary}.
Only the first line of Eq.~\eqref{eq:eq_system} is a system of \acp{ode}, in two variables; the other two lines consist of three
quadratures, taken afterwards. The explicit expressions for the equations
solved here are given in App.~\ref{app:higher}.

The complete workflow is as follows: 
(i) given some initial conditions in terms of the phase-space point $(r, p_{r_*}, \varphi, p_\varphi)$, compute the related slow
variables $\bar c_\mu$ through the inverse map of Eq.~\eqref{eq:cbar_of_c};
(ii) evolve the \ac{ode} system until a stopping condition is met. This typically happens when
an ``adiabaticity'' parameter, the fractional change of the binding energy over one radial period, crosses a
chosen threshold, or when the periastron $r_-$ falls below a minimum radius;
(iii) once the \ac{pa} stage reaches this condition (the handoff), the mean elements are mapped into the osculating ones, and then further into the phase-space point $(r, p_{r_*}, \varphi, p_\varphi)$;
(iv) this set of values is used to evolve the standard \acp{eom} until merger;
(v) the dynamics is reconstructed over the span covered by the \ac{pa} stage, from the initial time to the handoff: at each
sample the mean elements are mapped into the osculating ones, and then into $(r, p_{r_*}, \varphi, p_\varphi)$, on which the
(inspiral) waveform is computed. The samples are spaced by a fixed fraction of the orbital period, with $n_{\rm per}$ points per
period set by the initial eccentricity;
(vi) for \acp{bbh}, the attachment of the merger-ringdown proceeds as in standard \ac{eob} models.

\begin{figure}[t]
    \centering
    \includegraphics[width=\columnwidth]{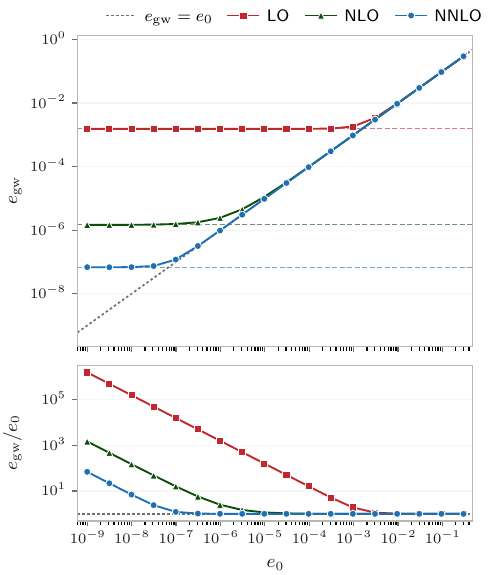}
    \caption{\label{fig:lowecc_ics}
    Eccentricity measured from the waveform vs initial parameter eccentricity
    in the low-eccentricity regime, for an equal-mass, nonspinning binary starting at $M f_0 = 0.003$. The different orders, shown
    in different colors, saturate at different eccentricity values, marked by the dashed lines; the bottom panel shows the ratio
    of the two eccentricities. Including corrections up to \ac{nnlo} improves the agreement with the gauge-invariant eccentricity
    measure down to $e_0 \simeq 7 \times 10^{-8}$.
    }
\end{figure}

\begin{figure*}[tp]
    \centering
    \includegraphics[width=0.49\textwidth,height=0.8\textheight,keepaspectratio]{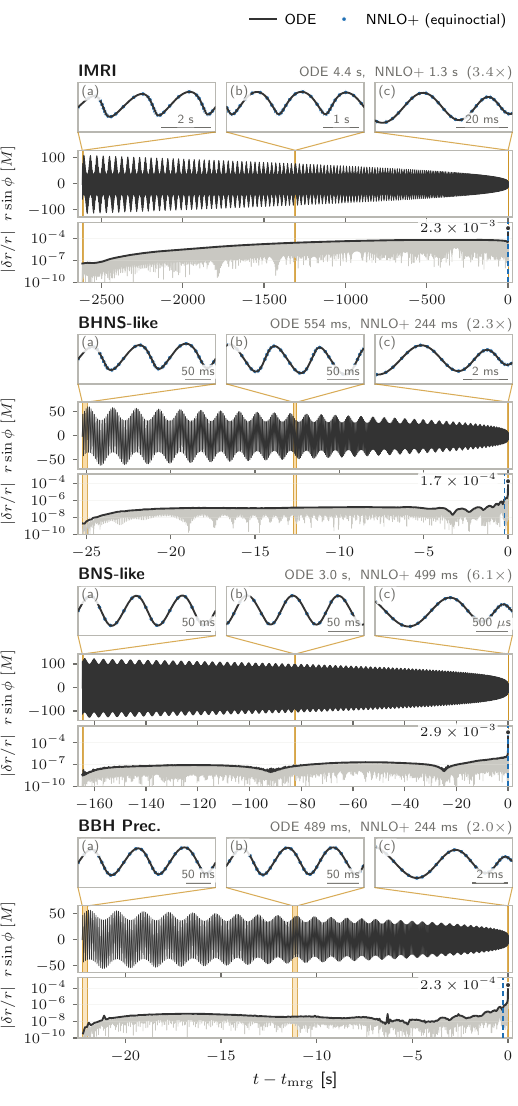}
    \includegraphics[width=0.49\textwidth,height=0.8\textheight,keepaspectratio]{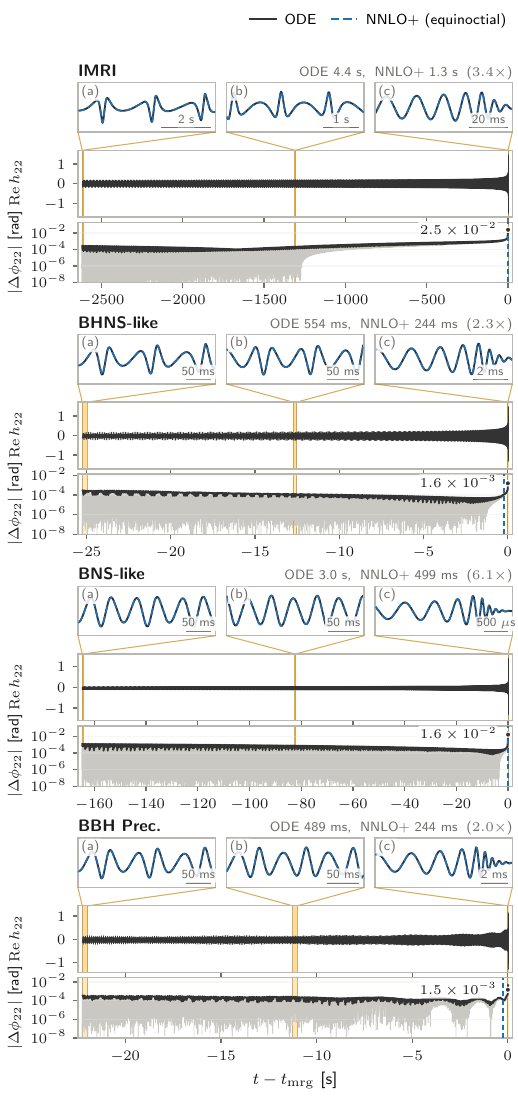}
    \caption{\label{fig:dynandwf}
    Dynamics (left) and $(2,2)$ waveform (right) for four representative systems:
    an intermediate mass-ratio inspiral (\acs{imri}; $M = 100 M_\odot$, $q = 20$, $\chi_1 = 0.5$, $\chi_2 = 0$, $e_0 = 0.49$ from $1$ Hz),
    a \ac{bhns}-like binary ($M = 10 M_\odot$, $q = 5$, $\chi_1 = 0.5$, $\chi_2 = -0.5$, $e_0 = 0.3$ from $20$ Hz), a
    \ac{bns}-like binary ($M = 2.7 M_\odot$, $q = 1$, nonspinning, $e_0 = 0.1$ from $20$ Hz, no tides) and a precessing
    \ac{bbh} ($M = 10 M_\odot$, $q = 3$, $\bm{\chi}_1 = (0.7, 0, 0.1)$, $\chi_2 = 0$, $e_0 = 0.2$ from $20$ Hz), for which the
    inertial-frame $(2,2)$ mode is shown; sampled at $1024$, $4096$, $8192$ and $4096$~Hz, respectively.
    The solid lines correspond to the \ac{ode} solution, while dashed lines denote the \ac{nnlo}+ solution.
    The bottom panels show the relative difference in the radius, $|\delta r/r|$ (left), and the phase difference of the
    $(2,2)$ mode, $|\Delta\phi_{22}|$ (right; inertial frame for the precessing binary); the dashed vertical line marks the handoff from
    the \ac{pa} stage to the \ac{ode}. The \ac{pa} is in excellent
    agreement with the full numerical \ac{ode}, with phase differences of the order of $10^{-4}$ rad up to the handoff and of at
    most $\PAvsODEDphiMergerImri$ rad at merger.
    }
\end{figure*}

\begin{figure*}[t]
    \centering
    \includegraphics[width=\textwidth]{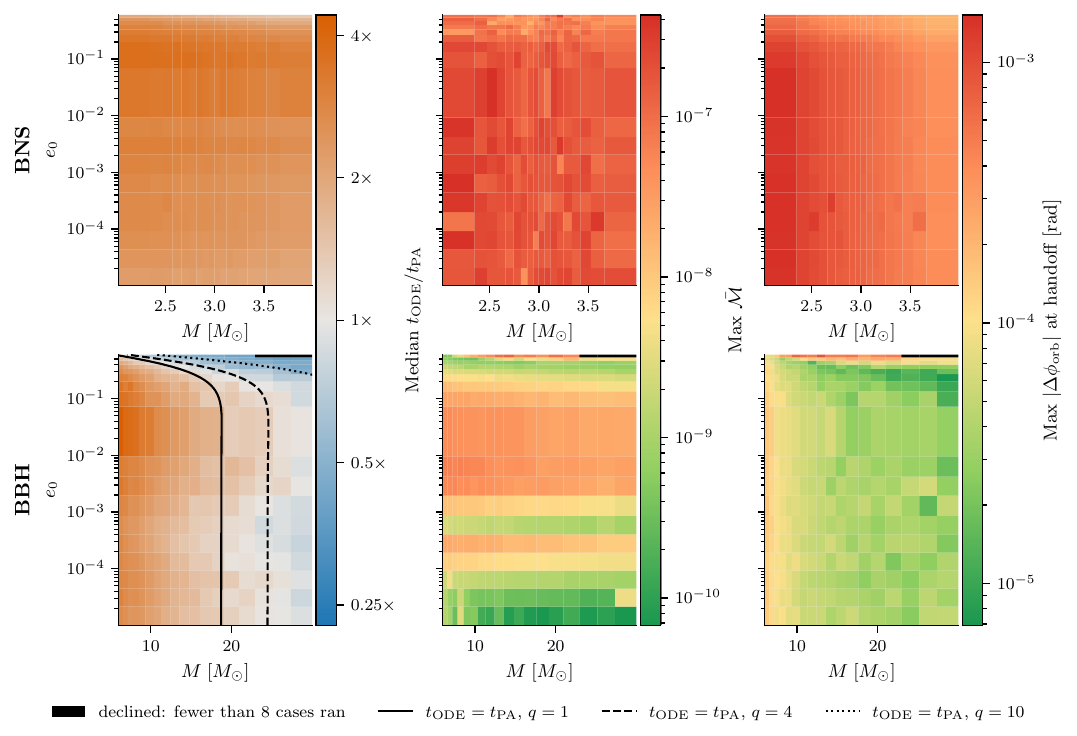}
    \caption{\label{fig:sweep} Comparison between the \ac{pa} and the \ac{ode} solutions over $10^4$ \acp{bns} (top) and $10^4$
    \acp{bbh} (bottom), as a function of the total mass and of the initial eccentricity at $20$ Hz, which is drawn half
    log-uniformly below $10^{-2}$ and half uniformly above. From left to right, in each cell: median speedup, largest mismatch, and largest
    orbital dephasing at the handoff without any time or phase alignment. Black cells are those in which the \ac{pa} stage ran for fewer than
    eight systems, the others falling back to the \ac{ode} because the adiabaticity threshold is crossed almost at once
    (App.~\ref{app:techimp}). The black lines in the speedup panels mark $t_{\rm ODE} = t_{\rm PA}$, as determined in
    Sec.~\ref{sec:benchmarks}, for mass ratios $q = 1$ (solid), $4$ (dashed) and $10$ (dotted); for \acp{bns} the $q = 1$ line
    lies beyond the masses shown.}
\end{figure*}

\begin{figure*}[t]
    \centering
    \includegraphics[width=\textwidth]{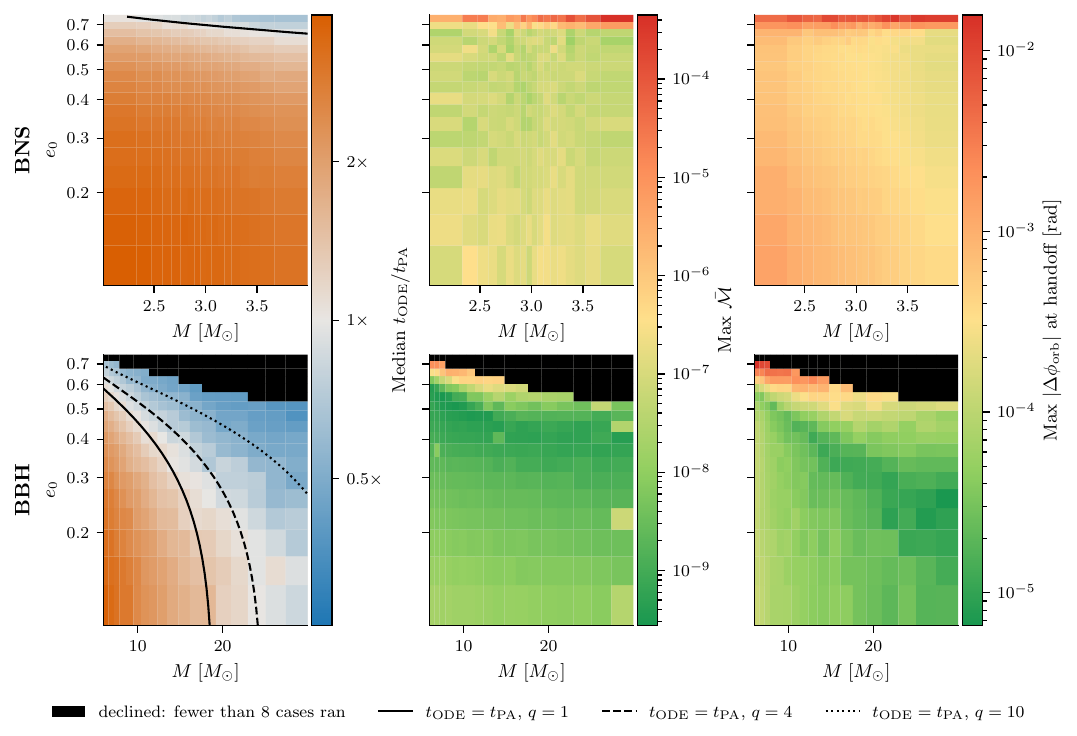}
    \caption{\label{fig:sweep_highecc} Same as Fig.~\ref{fig:sweep}, for the high-eccentricity sweep of $10^4$ \acp{bns} and $10^4$ \acp{bbh}: $e_0$ is drawn uniformly in
    $[0.1, 0.75]$, and up to $80$ points per orbital period are allowed.}
\end{figure*}

\section{Application to \dali{} and performance}
\label{sec:application}
After a brief aside dedicated to the description
of \dali{}, in this section I discuss the application of the \ac{pa} equinoctial technique described in the
previous section to the calculation of initial conditions and their subsequent evolution.
The highest level I implement is \ac{nnlo}+, i.e.\ the \ac{nnlo} level of Sec.~\ref{sec:higherpa} with the map from mean to
osculating elements carried to third order: $(3, 3, 2)$ instead of $(2, 3, 2)$. This is done
to retain a higher accuracy at large eccentricities.

\subsection{\dali{}: a brief overview}
\label{sec:dali}

\dali{} is a state-of-the-art \ac{eob} model designed for the accurate and efficient evaluation
of the dynamics and gravitational waveforms of generic binary systems~\cite{Chiaramello:2020ehz,Nagar:2021gss,Nagar:2024dzj,Nagar:2024oyk,Gamba:2024cvy,Albanesi:2025txj},
including tidal interactions~\cite{Damour:2009wj,Bernuzzi:2014owa,Akcay:2018yyh,Gamba:2022mgx,Gamba:2023mww}
and black hole--neutron star mergers~\cite{Gonzalez:2022prs,Gonzalez:2025xba}.
Written in \texttt{C99} with \texttt{python} bindings for integration into data-analysis
pipelines, its modular codebase allows for straightforward implementation of additional physics.
The core workflow follows the same rationale as the list presented
at the end of Sec.~\ref{sec:equinoct}: initialization, calculation of initial data,
integration of the dynamical equations, and waveform generation. For precessing binaries,
the spin dynamics and frame rotation (the ``twist''~\cite{Buonanno:2002fy,Buonanno:2005xu,Schmidt:2010it,Boyle:2011gg,OShaughnessy:2011pmr,Schmidt:2012rh}) are evolved separately
from the orbit, driven by its (orbital) frequency~\cite{Akcay:2020qrj,Gamba:2021ydi,Gamba:2024cvy}.
For \ac{fd} quasi-circular \ac{bns} waveforms, the modes and twist are then \ac{spa}-transformed,
leading directly to the computation of the final waveform polarizations.

Since the precessional dynamics are evaluated after the orbital evolution
and do not back-react on the orbital dynamics, we can apply the \ac{pa} framework
directly to our spin-precessing evolutions without structural modification.

Because the overall computational cost of any \ac{eob} model is dominated by the integration of its
governing \acp{ode}, we complement the \ac{pa} implementation with several
optimizations to the baseline solver:
\begin{itemize}\item
    We tailor the choice of \ac{ode} solver to the binary system,
    employing a multistep Adams method for \acp{bbh}, and an eighth-order Runge--Kutta
    Prince--Dormand method for systems with tides (\acp{bhns} and \acp{bns}).
    \item We optimize the absolute and relative error tolerances of the \ac{ode}
    solver of the spin dynamics.
    \item For precessing binaries, the rotation coefficients are built on a grid of their own, set by the evolution
    of the Euler angles, and interpolated onto the output samples (similarly to Ref.~\cite{Whittall:2026yvp}), and the
    polarizations are summed directly from the co-precessing modes when the inertial ones are not required~\cite{Ramos-Buades:2023ehm}.
    \item We cache repeatedly evaluated dynamical quantities (metric potentials,
    radiation reaction fluxes, the Hamiltonian itself, $\dots$) to eliminate redundant function evaluations.
    \item We introduce optional interpolation of selected dynamical quantities over Chebyshev
    grids to reduce computational overhead; it is active in all the runs of this work, with and without the \ac{pa} stage.
\end{itemize}
Together, these performance enhancements substantially improve the overall
efficiency of the model with a negligible impact on numerical accuracy.

Finally, the work presented here builds upon a development version of \dali{} that
delivers significantly improved accuracy across the parameter space~\cite{ChiaramelloInPrep,RettegnoInPrep,AlbanesiInPrep}.
The \ac{pa} stage, however, makes no assumption on the form of the \ac{eob} metric potentials
or of the radiation reaction: it evaluates whatever the model selects, and so applies unchanged to other versions of the model.
If the model were fixed once and for all, the orbit-averaged rates could instead be
tabulated offline as functions of the mean elements, as is standard in the \ac{gsf} community~\cite{Chua:2020stf,Katz:2021yft,Chapman-Bird:2025xtd,Lynch:2026ibo};
I prefer to keep the framework flexible, and leave this to future work.

\subsection{Initial conditions}
\label{sec:ics}
The calculation of accurate, representative initial data for the evolution of a desired two-body system is crucial,
as it directly affects the accuracy of the subsequent evolution and waveform.
Considerable care has been taken in the past in the construction of eccentricity-reduced initial
data~\cite{Pfeiffer:2007yz,Buonanno:2010yk,Ramos-Buades:2018azo}, where spurious oscillations in the radial separation between the two bodies are minimized.
The bridge between the high eccentricity regime and the $e=0$ limit
is still relatively unexplored, at least in the context of state-of-the-art \ac{eob} models.
Indeed, both the \texttt{SEOBNR} and \texttt{TEOBResumS} families currently contain -- at most -- corrections
to the 1\ac{pa} level~\cite{Ramos-Buades:2021adz,Gamboa:2024hli,Nagar:2024dzj,Gamboa:2026jht}, expressed as \ac{pn} expansions rather than derived from the underlying \ac{eob} Hamiltonian~\cite{Ramos-Buades:2021adz,Gamboa:2024imd}
or obtained via formally incorrect (but practically useful) procedures~\cite{Bonino:2022hkj}.
Here the goal is to construct initial data that accurately represents a desired eccentricity
in the $e\rightarrow 0$ limit to below the \ac{nr} error level, and that is entirely consistent with the underlying
\ac{eob} dynamics.

Following Refs.~\cite{Ramos-Buades:2023yhy,Gamboa:2024hli,Nagar:2024dzj}, the input parameters to the initial data construction are
$(e, f_0, \zeta)$, where $f_0$ is the orbit-averaged frequency of the (2,2) mode and 
$\zeta$ is the true anomaly measured on the mean orbit identified by $e = \bar r_a / r_c$.
Then, $\bar p_\varphi$ and $\bar r_a = |\bar z|$ are fixed by two conditions:
\begin{equation}
    \label{eq:ic_pair}
    \bar r_a = e\, r_c(\bar p_\varphi, \bar r_a) , \qquad (1+k)\,n = \pi f_0 .
\end{equation}

The starting anomaly $\zeta$ is defined through $r = p/(1 + e\cos\zeta)$, with $p = r_c (1 - e^2)$ the semi-latus rectum
of the mean orbit, so that $r$ spans its turning points; $\zeta$ is
therefore the true anomaly in the Newtonian limit, while $X$ is an eccentric-like anomaly.
Setting $\bar w = 0$ at the start, the corresponding anomaly on
the mean orbit, $X_{\rm m}$, is given by
\begin{equation}
    \label{eq:ic_anomaly}
    \cos X_{\rm m} = \frac{\cos\zeta + e}{1 + e \cos\zeta} ,
\end{equation}
with $\sin X_{\rm m}$ of the same sign as $\sin\zeta$, and the mean longitude follows from the first of Eqs.~\eqref{eq:TV_maps},
\begin{equation}
    \label{eq:Xm}
    \bar\Lambda = X_{\rm m} + T(X_{\rm m}; \bar c_\mu) .
\end{equation}
The osculating elements are then obtained via:
\begin{equation}
    \label{eq:ic_map}
    \begin{aligned}
    p_\varphi &= \bar p_\varphi + \epsilon\, \tilde{c}^{(1)}_{p_\varphi} + \epsilon^2\, \tilde{c}^{(2)}_{p_\varphi} , \\
    z         &= \bar z + \epsilon\, \tilde{c}^{(1)}_z + \epsilon^2\, \tilde{c}^{(2)}_z , \\
    \Lambda   &= \bar\Lambda + \epsilon\, \tilde\Lambda^{(1)} + \epsilon^2\, \tilde\Lambda^{(2)} ,
    \end{aligned}
\end{equation}
where the $\tilde\Lambda^{(i)}$ are the oscillating corrections to the mean longitude introduced in
Sec.~\ref{sec:equinoct}. Finally, the anomaly $X$ on the osculating orbit is the solution of
\begin{equation}
    X + T(X; c_\mu) = \Lambda ,
\end{equation}
and the initial phase-space point follows from Eqs.~\eqref{eq:eq_param} and~\eqref{eq:prstar_of_X}, with $\varphi_0 = 0$.

Figure~\ref{fig:lowecc_ics} shows the effect of different \ac{pa} orders on the initial eccentricity estimated
from the waveform using the \texttt{gw\_eccentricity} library~\cite{Shaikh:2023ypz,Shaikh:2025tae} for a reference equal-mass binary.
While \ac{lo} initial conditions leave a spurious eccentricity of $1.5\times10^{-3}$ in the early inspiral, the inclusion of \ac{nlo} and \ac{nnlo} corrections
significantly lowers this floor, to $1.5\times10^{-6}$ and $7\times10^{-8}$, respectively. This value is well below the residual
eccentricity of eccentricity-reduced \ac{nr} simulations, which reaches $10^{-4}$--$10^{-5}$~\cite{Pfeiffer:2007yz,Buonanno:2010yk,Ramos-Buades:2018azo,Boyle:2019kee}.

\subsection{Benchmarks}
\label{sec:benchmarks}
Let us now turn to benchmarking the equinoctial \ac{pa} framework in terms of accuracy and speed.
A priori, this method is expected to be most useful for those systems
for which the native \ac{ode} solver of the model takes longest, and whose evolution is at the same time close to adiabatic.
These are, for instance, \ac{bns}-like systems with low total masses; systems with large mass ratios
like \ac{bhns}; or \ac{bbh} systems evolving from very low initial frequencies,
like \acp{imri} in the bands of the Einstein Telescope~\cite{Punturo:2010zz} and Cosmic Explorer~\cite{Reitze:2019iox}.

Representative comparisons for these three cases, and for a precessing \ac{bbh}, are shown in Fig.~\ref{fig:dynandwf} for
\ac{nnlo}+~\ac{pa} evolutions\footnote{The timings quoted here and in the figure are medians of five runs on a laptop.}.
Here the \ac{pa} framework delivers considerable speedups with respect to the baseline \ac{ode} solver, at negligible loss of accuracy: $\times \PAvsODESpeedupImri$ for the \ac{imri},
$\times \PAvsODESpeedupBhnsLike$ for the \ac{bhns}-like binary, $\times \PAvsODESpeedupBnsLike$ for the \ac{bns}-like one and
$\times \PAvsODESpeedupBbhPrec$ for the precessing \ac{bbh}. At the handoff, the relative error on the radial separation (the
largest value over each cycle, as drawn in the bottom panels of Fig.~\ref{fig:dynandwf}) is $\PAvsODEDrHandoffImri$ for the
\ac{imri}, which is also the most eccentric, and $\PAvsODEDrHandoffBhnsLike$, $\PAvsODEDrHandoffBnsLike$ and
$\PAvsODEDrHandoffBbhPrec$ for the other three, and the waveform phase error $\Delta\phi_{22}$ is $\PAvsODEDphiHandoffImri$,
$\PAvsODEDphiHandoffBhnsLike$, $\PAvsODEDphiHandoffBnsLike$ and $\PAvsODEDphiHandoffBbhPrec$ rad, in the same order.
Both grow after the handoff, over the last $\sim 20$ orbits, where the two runs are integrated with the \ac{ode}: at merger the
error on the radial separation is $\PAvsODEDrMergerImri$, $\PAvsODEDrMergerBhnsLike$, $\PAvsODEDrMergerBnsLike$ and
$\PAvsODEDrMergerBbhPrec$, and $\Delta\phi_{22}$ is $\PAvsODEDphiMergerImri$, $\PAvsODEDphiMergerBhnsLike$,
$\PAvsODEDphiMergerBnsLike$ and $\PAvsODEDphiMergerBbhPrec$ rad. For the \ac{bns}-like
system, waveform generation falls below one second.

A more careful profiling of the code reveals that the cost of the \ac{pa} framework is split between two parts: the secular
evolution, whose cost scales with the number of secular steps, and the reconstruction of the dynamics together with the
computation of the waveform on its samples, whose cost scales with the number of orbits. For short signals the former
dominates: for a \ac{bbh} with $M = 27 M_\odot$, $q = 3$ and $e_0 = 0.1$ it takes about $70\%$ of the run, mostly in building
the second-order corrections and in the numerical inversions they require (see App.~\ref{app:techimp}), while the reconstruction and
the waveform take a few percent. As the number of orbits grows, the latter become increasingly important: for the \ac{imri},
\ac{bhns}-like and \ac{bns}-like systems of Fig.~\ref{fig:dynandwf}, the secular evolution takes $21\%$, $55\%$ and
$26\%$ of the run, and the reconstruction and the waveform $45\%$, $19\%$ and $36\%$, respectively. 

To investigate the performance of the \ac{pa} model more systematically, I consider $10^4$ \acp{bbh}
and $10^4$ \acp{bns}, with total masses $M_{\rm BBH} \in [6, 30] M_\odot$, distributed uniformly in $\log M$
(component masses $m_{1,2} \in [1, 2] M_\odot$ for \acp{bns}, i.e. $M_{\rm BNS} \in [2, 4] M_\odot$),
mass ratios $q \in [1, 10]$, capped so that the lighter black hole has at least $2.5 M_\odot$ (for \acp{bns}, as fixed by the
component masses), spins $\chi_{1,2} \in [-0.9, 0.9]$ ($[-0.1, 0.1]$) and, for \acp{bns}, tidal parameters
$\Lambda_{1,2}$ obtained by solving the Tolman-Oppenheimer-Volkoff equations for one of five equations of state, drawn at
random for each system: SLy4~\cite{Douchin:2001sv}, ENG~\cite{Engvik:1995gn}, MPA1~\cite{Muther:1987xaa}, H4~\cite{Lackey:2005tk}
and MS1b~\cite{Mueller:1996pm}. The initial eccentricity at $20$ Hz is drawn half uniformly in $\log e_0$ within
$[10^{-5}, 10^{-2}]$ and half uniformly in $e_0$ within $[10^{-2}, 0.6]$; a second parameter space exploration,
of again $10^4$ \acp{bbh} and $10^4$ \acp{bns} with $e_0$ drawn uniformly in $[0.1, 0.75]$, probes high eccentricities
with a denser sampling of the orbital period ($n_{\rm per} \leq 80$, see App.~\ref{app:techimp}).
Note that we focus here on aligned-spin systems, since precessing binaries are in any case obtained by twisting-up the
aligned-spin modes. The same parameter spaces with precessing spins are explored in App.~\ref{app:precessing}, with
similar results.

For each system, I time waveform generation with the \ac{ode} and with the \ac{pa} framework, and quote medians over
the systems that fall in the same cell of the parameter space, about $30$ per cell on average.
The two runs of each system are timed back to back on the same machine.
I compute the median speedup $t_{\rm ODE}/t_{\rm PA}$, as well as the mismatch $\bar{\mathcal{M}}$.
The latter is a measure of the disagreement between the two waveforms $h_{\rm PA}$ and $h_{\rm ODE}$, and is defined as:
\begin{equation}
    \label{eq:mismatch}
    \bar{\mathcal{M}} = 1 - \max_{\Delta t, \Delta \varphi_0} \frac{(h_{\rm ODE}, h_{\rm PA})}{\sqrt{(h_{\rm ODE}, h_{\rm ODE}) (h_{\rm PA}, h_{\rm PA})}} \, ,
\end{equation}
where:
\begin{equation}
    (h_1, h_2) = 4 \Re \int \frac{\tilde{h}_1 \tilde{h}_2^*}{S_n(f)} df \, .
\end{equation}
Since the aim is to compare two solutions of the same model rather than to assess detectability, the mismatch is computed with
\texttt{PyART} and \texttt{pycbc}~\cite{PyART,pycbc} on the $(2,2)$ mode with a flat spectrum, $S_n(f) = 1$, between $1.25 f_0$ and $0.45$ times the
sampling rate after cutting the two waveforms to their common length. The waveforms are sampled at $16384$~Hz for
$M < 8 M_\odot$, i.e.\ for all the \acp{bns} and the lightest \acp{bbh}, and at $8192$~Hz above.
For reference, mismatches between semi-analytical waveform models and \ac{nr} are $\mathcal{O}(10^{-5})$ in optimistic
scenarios, and $\mathcal{O}(10^{-4})$--$\mathcal{O}(10^{-3})$ on average~\cite{Pompili:2023tna,Nagar:2023zxh,Nagar:2024oyk,Gamboa:2024hli,Gamboa:2026jht}.

Figure~\ref{fig:sweep} summarizes the results for the low-eccentricity parameter space
and Fig.~\ref{fig:sweep_highecc} those for the high-eccentricity one.
For $e_0 \leq 0.6$ the median mismatch is $\SwMmMedBbhLow$ for \acp{bbh} and $\SwMmMedBnsLow$ for \acp{bns},
and the largest values found are $\mathcal{O}(10^{-7})$ for both, with differences between \ac{ode} and \ac{pa}
remaining small also in the circular limit as expected.
At this level of accuracy, the mismatch of the \acp{bbh} is set by the sampling of the reconstructed dynamics: at fixed
$n_{\rm per}$ it grows with $e_0$, and it drops each time $n_{\rm per}$ increases, which produces the horizontal bands of
Fig.~\ref{fig:sweep}. 
In terms of the orbital phases, the difference at the handoff is at most $\SwDphiHandoffMaxBbh$ rad for \acp{bbh} and
$\SwDphiHandoffMaxBns$ rad for \acp{bns}, over inspirals that accumulate up to $\mathcal{O}(10^{4})$ rad.
At large eccentricities the accuracy degrades, although not dramatically. For $0.6 < e_0 \leq 0.7$ the median mismatch is
$\SwMmMedBbhSixSeven$ for \acp{bbh}, while for \acp{bns} it is $\SwMmMedBnsSixSeven$. Above
$e_0 = 0.7$ the medians are $\SwMmMedBbhAboveSeven$ and $\SwMmMedBnsAboveSeven$. Only $\SwMmNBig$ systems out of $2 \times 10^4$, all with
$e_0 > \SwMmBigEmin$, have $\bar{\mathcal{M}} > 10^{-4}$. 
The reason for this degradation is that as the eccentricity increases the system becomes less adiabatic, and the
assumptions underlying the \ac{pa} stage less accurate.

Let us now turn to the speed. The gain is determined by how long the inspiral is, and therefore by the total mass and, to a lesser
extent, by the eccentricity. For \acp{bns} the median speedup over the low-eccentricity parameter space is $\sim \SwSpeedupBnsLow$, with $90\%$ of the
systems between $\SwSpeedupBnsLowPfive$ and $\SwSpeedupBnsLowPninetyfive$; it is largest, about $\SwSpeedupBnsLowEcc$, for $0.01 < e_0 \leq 0.2$, and decreases to $\SwSpeedupBnsFiveSix$ for $0.5 < e_0 \leq 0.6$ and to $\SwSpeedupBnsAboveSeven$ above
$e_0 = 0.7$. The reason is twofold: the number of nodes of the orbit averages (App.~\ref{app:techimp}) and the number of points per orbital
period needed to resolve the orbit grow, while the evolution becomes shorter, reducing the overall speedup.
In absolute terms, the median time needed to generate a \ac{bns} waveform from $20$ Hz goes from $\sim \SwTimeODEBns$ s to $\sim \SwTimePABns$ s, and $\SwFracBnsSubsecond\%$ of the systems take less
than one second ($\SwFracBnsSubsecondLowEcc\%$ of those with $e_0 < 0.3$).

For \acp{bbh} the picture is different. The median speedup is $\sim \SwSpeedupBbhLight$ for
$M \leq 8 M_\odot$ and decreases steadily with the total mass, crossing unity at $M \simeq \SwMassBbhUnity M_\odot$ and reaching $\SwSpeedupBbhHeaviest$ at $M = 30 M_\odot$.
Similarly, at fixed mass it decreases with the eccentricity, and its median falls below unity for $e_0 \gtrsim \SwEccBbhUnity$.
Overall the \ac{pa} stage is slower than the \ac{ode} integration for $\SwFracBbhSlower\%$ of the \acp{bbh} of the low-eccentricity parameter space.
These are, however, systems whose waveform is generated in $\sim \SwTimeBbhSlowerLo$--$\SwTimeBbhSlowerHi$ s in either case.

The parameter space explorations also allow one to determine the regions
where the \ac{pa} stage is more efficient than the \ac{ode}, given by the condition $t_{\rm ODE}/t_{\rm PA} = 1$.
At fixed eccentricity, the gain is approximately set by the number of \ac{gw} cycles, which
is then the natural coordinate to describe the speedup.
At leading post-Newtonian order for a circular binary, this is
\begin{equation}
    \label{eq:ncyc}
    N_{\rm cyc} = \frac{1}{32\pi\nu} \left( \frac{\pi G M f_0}{c^3} \right)^{-5/3} .
\end{equation}
Fitting the logarithm of $t_{\rm ODE}/t_{\rm PA}$ over the full parameter space, excluding the cases in which the \ac{pa} stage does not run and the
outliers beyond three standard deviations, with a cubic polynomial in $x = \log_{10} N_{\rm cyc}$ and $e_0$, one finds:
\begin{equation}
    \label{eq:speedup_fit}
    \ln \frac{t_{\rm ODE}}{t_{\rm PA}} \simeq \sum_{i + j \leq 3} a_{ij}\, x^i e_0^j ,
\end{equation}
with the coefficients $a_{ij}$ of Table~\ref{tab:speedup_fit}.

\begin{table}[t]
    \caption{\label{tab:speedup_fit} Coefficients $a_{ij}$ of the fit of Eq.~\eqref{eq:speedup_fit}, with
    $x = \log_{10} N_{\rm cyc}$. The terms cancel strongly, and three significant figures are needed to keep
    $N_{\rm crit}$ within $1\%$.}
    \begin{ruledtabular}
    \begin{tabular}{crrrr}
          & $e_0^0$  & $e_0^1$  & $e_0^2$  & $e_0^3$ \\
    \hline
    $x^0$ & $\SwFitaa$ & $\SwFitab$ & $\SwFitac$ & $\SwFitad$ \\
    $x^1$ & $\SwFitba$ & $\SwFitbb$ & $\SwFitbc$ &            \\
    $x^2$ & $\SwFitca$ & $\SwFitcb$ &            &            \\
    $x^3$ & $\SwFitda$ &            &            &            \\
    \end{tabular}
    \end{ruledtabular}
\end{table}

The fit is only valid over the range covered by the sweeps (their 1st to 99th percentiles), $\SwFitNcycLo \lesssim N_{\rm cyc} \lesssim \SwFitNcycHi$
and $e_0 \leq 0.75$, and
should not be extrapolated beyond it. Its zero defines the boundary $N_{\rm crit}(e_0)$, drawn as black
lines in Figs.~\ref{fig:sweep} and~\ref{fig:sweep_highecc}. 
The criterion $N_{\rm cyc} > N_{\rm crit}(e_0)$ selects the faster option for 
$\SwFitRightBbh\%$ of the \acp{bbh} and $\SwFitRightBns\%$ of the \acp{bns}.

\subsection{Comparisons with other models}
\label{sec:comparison}

We now briefly compare the performance of \dali{} with \texttt{SEOBNRv6EPHM}, \texttt{SEOBNRv5PHM}, and with \dali{} itself
in its previous version (v1.1.2). Since the latter two are also benchmarked in Ref.~\cite{Gamboa:2026ybr},
they allow us to relate our timings to those of \texttt{SEOBNRv6EPHM}. 
Other models for eccentric, precessing binaries exist, such as the \ac{eob} model
\texttt{SEOBNRE}~\cite{Cao:2017ndf,Liu:2021pkr,Liu:2023ldr} and the inspiral-only \ac{pn} models
\texttt{pyEFPEHM}~\cite{Klein:2018ybm,Klein:2021jtd,Morras:2025nlp,Morras:2026fho}, but here we
restrict to the \texttt{SEOBNR} family, which, like \dali{}, models the full inspiral-merger-ringdown, and has been used to analyse observed events.
Since \texttt{SEOBNRv6EPHM} is not yet publicly available 
, we perform the comparison by relying on the data behind Fig.~11 of Ref.~\cite{Gamboa:2026ybr}.
We consider four representative systems among the events detected by Advanced LIGO~\cite{LIGOScientific:2014pky},
Advanced Virgo~\cite{VIRGO:2014yos} and KAGRA~\cite{KAGRA:2020tym},
with parameters drawn from the posterior samples of \texttt{SEOBNRv6EPHM}~\cite{Gamboa:2026zenodo} for GW150914~\cite{LIGOScientific:2016aoc},
GW190521~\cite{LIGOScientific:2020iuh}, GW200105~\cite{LIGOScientific:2021qlt} and GW170817~\cite{LIGOScientific:2017vwq}\footnote{The
\texttt{SEOBNRv6EPHM} model does not provide a description of tidal effects on eccentric orbits yet,
so all binaries are actually \acp{bbh}}.
We draw 128 posterior samples for each system and generate the corresponding
waveforms with \dali{}, \dali{} v1.1.2 and \texttt{SEOBNRv5PHM}~\cite{Ramos-Buades:2023ehm,Estelles:2025zah} (from the \texttt{pyseobnr} 0.2.7 and 0.3.7 releases).
Every model is called through the \texttt{gwsignal} interface of
\texttt{lalsuite}~\cite{lalsuite}, the way \texttt{bilby}~\cite{Ashton:2018jfp,Romero-Shaw:2020owr} would in a parameter-estimation run,
so that each evaluation also includes conditioning and \ac{fft}-ing\footnote{For \dali{}, the waveform starts at the analysis'
minimum frequency, its start is tapered, and it is placed in a segment of the analysis' duration and transformed with an \ac{fft};
\texttt{pyseobnr} conditions its own output.}.
For each model we first perform three evaluations, which we discard
so that one-off setup costs do not enter the timings, and then time one evaluation per sample.
For \dali{} we show for each event the faster of the \ac{ode} and \ac{pa} evolutions; this is also the choice made by the criterion
$N_{\rm cyc} > N_{\rm crit}(e_0)$ of Sec.~\ref{sec:benchmarks} for every sample (\ac{ode} for GW150914 and GW190521,
\ac{pa} for GW200105 and GW170817).
For the latter two, the optimized \ac{ode} evolution alone would be $\VcmpOdeOverPaMin$--$\VcmpOdeOverPaMax$ times slower.

\begin{figure}[t]
    \centering
    \includegraphics[width=\columnwidth]{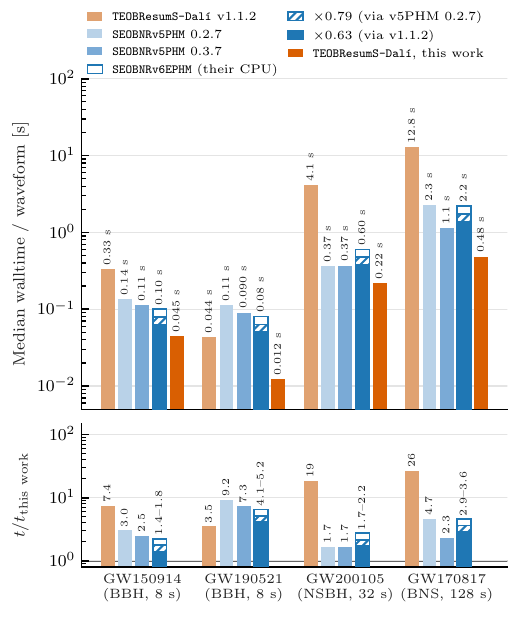}
    \caption{\label{fig:timing_v6cmp}
    Median walltime per waveform evaluation for 128 posterior samples of four GW events, drawn from the
    \texttt{SEOBNRv6EPHM} posteriors of Ref.~\cite{Gamboa:2026ybr} and generated with the settings of the
    corresponding analyses, including conditioning and \ac{fft}. \dali{} uses the modes of
    \texttt{SEOBNRv6EPHM} and, for each event, the faster of the \ac{ode} and \ac{pa} evolutions; tides are
    switched off for GW170817, as in Ref.~\cite{Gamboa:2026ybr}.
    \dali{}, \dali{} v1.1.2 and \texttt{SEOBNRv5PHM} (\texttt{pyseobnr} 0.2.7 and 0.3.7; 
	these two with their default settings) are timed on a single core of an \VcmpCPU{} (\VcmpSystem). 
	\texttt{SEOBNRv6EPHM} is taken from Fig.~11 of Ref.~\cite{Gamboa:2026ybr}
    (outlined bar), and presented together with two estimates of its timings
    on our CPU, obtained by rescaling it with the mean ratio of our timings to those of Ref.~\cite{Gamboa:2026ybr}:
    $\VcmpFactorVfive$ for \texttt{SEOBNRv5PHM} 0.2.7 (hatched) and $\VcmpFactorVone$ for \dali{} v1.1.2, GW190521 excluded (solid).
    Bottom: ratio to \dali{}.}
\end{figure}

Results are presented in Fig.~\ref{fig:timing_v6cmp}. The bottom panel shows the ratio of each walltime to that of \dali{}:
$\VcmpVsVsixLo$--$\VcmpVsVsixHi$ for \texttt{SEOBNRv6EPHM}, $\VcmpVsVfiveMin$--$\VcmpVsVfiveMax$ for \texttt{SEOBNRv5PHM}, 
and $\VcmpVsVoneMin$--$\VcmpVsVoneMax$ for \dali{} v1.1.2.
For GW190521, the shortest signal, the \texttt{SEOBNR} models take about $80\%$ as long as for GW150914, while \dali{} takes
$\sim 3.5$ times less: it is possible that, for such short signals, the \texttt{SEOBNR} timings are dominated by fixed costs that do not depend
on the length of the signal (e.g., setting up the model and its initial conditions, attaching the merger-ringdown,
or the overhead of the \texttt{Python} layer) rather than by the evolution of the binary.

\section{Frequency-domain waveforms}
\label{sec:fd}
Since the secular evolution is smooth and monotonic, the \ac{pa} stage lends itself naturally to the direct computation of
\ac{fd} waveforms. The \ac{spa}, however, cannot be applied
to an eccentric multipole as a whole: its frequency oscillates over each radial period, so that the same frequency is reached
several times per orbit.
The solution is to write each multipole as a sum of chirps, one per radial harmonic.
This decomposition is by no means new: it underlies the \ac{dgi} parameterization~\cite{Damour:2004bz,Konigsdorffer:2006zt},
the eccentric
\ac{pn}~\cite{Peters:1963ux,Yunes:2009yz,Tessmer:2010sh,Mikoczi:2012qy,Klein:2018ybm,Klein:2021jtd,Morras:2025nlp,Morras:2026fho,Moore:2019xkm,Boetzel:2019nfw}
and \texttt{IMRPhenomXE}~\cite{Ramos-Buades:2026kbq} models, two-timescale self-force
waveforms~\cite{Hinderer:2008dm,Pound:2021qin} and the multivoice waveforms of
\texttt{FastEMRIWaveforms}~\cite{Chua:2020stf,Katz:2021yft,Chapman-Bird:2025xtd,Hughes:2021exa,Speri:2023jte}, and it has also been applied to
\ac{eob} waveforms~\cite{Hinderer:2017jcs,Patterson:2024vbo,Islam:2025rjl}.
Here, I adopt and apply it to \dali{}, making use of the equinoctial parameterization and the model's own multipoles.

\subsection{Harmonic decomposition and stationary-phase approximation}
\label{sec:fd_method}
At fixed elements, the quantity
\begin{equation}
h_{\ell m}\, e^{im\lambda} = \hat h_{\ell m}\, e^{-imV}
\end{equation}
is a periodic function of $X$ (and, as such, of $\Lambda$ and $\bar\Lambda$ as well), since
$\hat h_{\ell m} \equiv h_{\ell m}\, e^{im\varphi}$ depends on $(r, p_{r_*}, p_\varphi)$ alone.

It can therefore be Fourier decomposed as:
\begin{equation}
    \label{eq:fd_modes}
    h_{\ell m} = \sum_{j=-\infty}^{\infty} \mathcal{H}_{\ell mj}(\bar c_\mu)\, e^{-i\Psi_{mj}} , \qquad
    \Psi_{mj} \equiv m\bar\lambda + j\bar\Lambda .
\end{equation}
Given the rates $\dot{\bar\Lambda} = n + \bar F_\Lambda$ and $\dot{\bar\lambda} = (1+k)\,n + \bar F_\lambda$ of
Eq.~\eqref{eq:eq_system}, the $j$-th harmonic has frequency $\dot\Psi_{mj}/2\pi \simeq [m(1+k) + j]\, n/2\pi$.

To first order in $\epsilon$, i.e.\ including the displacement of the binary from the mean orbit, the
coefficients read (cf.\ the post-adiabatic corrections of Refs.~\cite{Damour:2004bz,Konigsdorffer:2006zt,Boetzel:2019nfw})
\begin{equation}
    \label{eq:fd_amp}
    \begin{aligned}
    \mathcal{H}_{\ell mj} &= \mathcal{H}^{(0)}_{\ell mj}
      + \epsilon \sum_{j'} \Bigl[ \frac{\partial \mathcal{H}^{(0)}_{\ell mj'}}{\partial \bar c_\mu}
        \tilde c^{(1)}_{\mu, j' - j} \\
    &\quad - i\, \mathcal{H}^{(0)}_{\ell mj'} \bigl( j'\, \tilde\Lambda^{(1)}_{j' - j}
        + m\, \tilde\lambda^{(1)}_{j' - j} \bigr) \Bigr] + O(\epsilon^2) , \\
    \mathcal{H}^{(0)}_{\ell mj} &= \frac{1}{N} \sum_{i=0}^{N-1} \hat h_{\ell m}(X_i)\, e^{-imV(X_i)} \\
    &\qquad \times e^{ij[X_i + T(X_i)]}\, n\, \frac{dt}{dX}\Big|_{X_i} ,
    \end{aligned}
\end{equation}
where the $X_i$ are the nodes used for the orbit averages, $\mathcal{H}^{(0)}_{\ell mj}$ are the coefficients on the mean
orbit, and the harmonics of the oscillating parts are defined by
\begin{equation}
\tilde c^{(1)}_\mu = \sum_p \tilde c^{(1)}_{\mu, p}\, e^{ip\bar\Lambda}
\end{equation}
and likewise for $\tilde\Lambda^{(1)}$ and $\tilde\lambda^{(1)}$. 
The derivatives of $\mathcal{H}^{(0)}_{\ell mj}$ are computed via finite differences.
The $N$ nodes resolve the harmonics with $|j| < N/2$; of these, I keep those whose amplitude exceeds $10^{-6}\,\|\mathcal{H}\|$,
with $\|\mathcal{H}\|^2 = \sum_j |\mathcal{H}_{\ell mj}|^2$, at some point of the secular evolution.
For the systems of Sec.~\ref{sec:fd_examples} this retains from $4$ harmonics (the quasi-circular \ac{bns}) to $49$
(the \ac{bbh} with $e_0 = 0.5$).

Efficient \ac{fd} representations of the $h_{\ell m}$ can then be obtained by
directly applying the \ac{spa} to Eq.~\eqref{eq:fd_modes}. This gives:
\begin{equation}
    \label{eq:fd_spa}
    \begin{aligned}
    \tilde h_{\ell m}(f) &\simeq \sum_j \mathcal{H}_{\ell mj}(t_j) \sqrt{\frac{2\pi}{i\,\ddot\Psi_{mj}(t_j)}}\;
                            e^{i[2\pi f t_j - \Psi_{mj}(t_j)]} , \\
    \dot\Psi_{mj}(t_j) &= 2\pi f .
    \end{aligned}
\end{equation}
As in Eq.~(421) of Ref.~\cite{Pound:2021qin} and in Refs.~\cite{Speri:2023jte,Ramos-Buades:2026kbq}, the stationary time $t_j$
is found by interpolation on the secular grid in $\bar p_\varphi$, without having to reconstruct the orbit, build a time grid or
take an \ac{fft}; the \ac{fd} phase is then the \ac{nnlo}+ secular phase itself.

One important caveat should be mentioned: while for $m>0$ modes all harmonics with $j \geq 0$ monotonically
increase in frequency, for some values of $j < 0$ zero-crossings and non-monotonic behavior can occur.
This is due to the growth of the periastron advance $k$, and causes
the direct \ac{spa} of Eq.~\eqref{eq:fd_spa} to fail.
The first value to exhibit this behavior is $j = -m - 1$,
which for $m=2$ corresponds to $\dot{\Psi}_{2,-3} = (2k - 1)\, n$.
Since however these harmonics reach lower frequencies than the
dominant ones, I directly transform them with an \ac{fft} on a coarse grid without
excessive computational cost. The grid spacing is given by $K/f_s$, where $f_s$ is the
sampling rate of the reference grid,
\begin{equation}
    \label{eq:fd_slow}
    K = 2^{\bigl\lfloor \log_2 [ f_s / (4 f_{\rm slow}) ] \bigr\rfloor} ,
\end{equation}
and $f_{\rm slow}$ is the largest frequency these harmonics reach.

\subsection{Joining the \texorpdfstring{\ac{spa}}{SPA} and the \texorpdfstring{\ac{fft}}{FFT}}
\label{sec:fd_join}
At late times, after the handoff, the \ac{spa} loses accuracy.
Since the \ac{ft} is linear, one can split the signal exactly as\footnote{
\texttt{SEOBNRv5THM}~\cite{Haberland:2025luz,Haberland:2026xvj} also combines the \ac{spa} and an \ac{fft}, but joins them
differently, mode by mode at a transition time and frequency.
}
\begin{equation}
    \label{eq:fd_split}
    \tilde h = \mathrm{FT}\bigl[(1 - \Theta)\, h\bigr] + \mathrm{FT}\bigl[\Theta\, h\bigr] ,
\end{equation}
with $\Theta(t)$ a smooth (Planck-taper) step that rises to one at the handoff.
The first term is given by Eq.~\eqref{eq:fd_spa} with the window evaluated at the stationary time of
each harmonic~\cite{Marsat:2018oam}, i.e.\ with amplitudes $[1 - \Theta(t_j)]\,\mathcal{H}_{\ell mj}(t_j)$.
For $j \geq 0$ I also add the next-order \ac{spa} terms in the derivatives of the window (Eq.~(51) of
Ref.~\cite{Marsat:2018oam}, with $[1 - \Theta]\, \mathcal{H}_{\ell mj}$ in place of $F$): they are small, and improve the agreement
between the \ac{spa} and the full \ac{fft} result. For $j < 0$ they are not small,
and they spoil the agreement with the full \ac{fft} result.
The second term is simply the \ac{fft} of the model's own \ac{td} waveform from the start of the window, which is short: the handoff
occurs $0.07$~s, or about $35$ \ac{gw} cycles, before merger for the \acp{bns} of Sec.~\ref{sec:fd_examples}, and $0.24$~s and $0.42$~s
before it for the two \acp{bbh}.

\subsection{Examples}
\label{sec:fd_examples}
\begin{figure*}[tp]
    \centering
    \includegraphics[width=\textwidth,height=0.9\textheight,keepaspectratio]{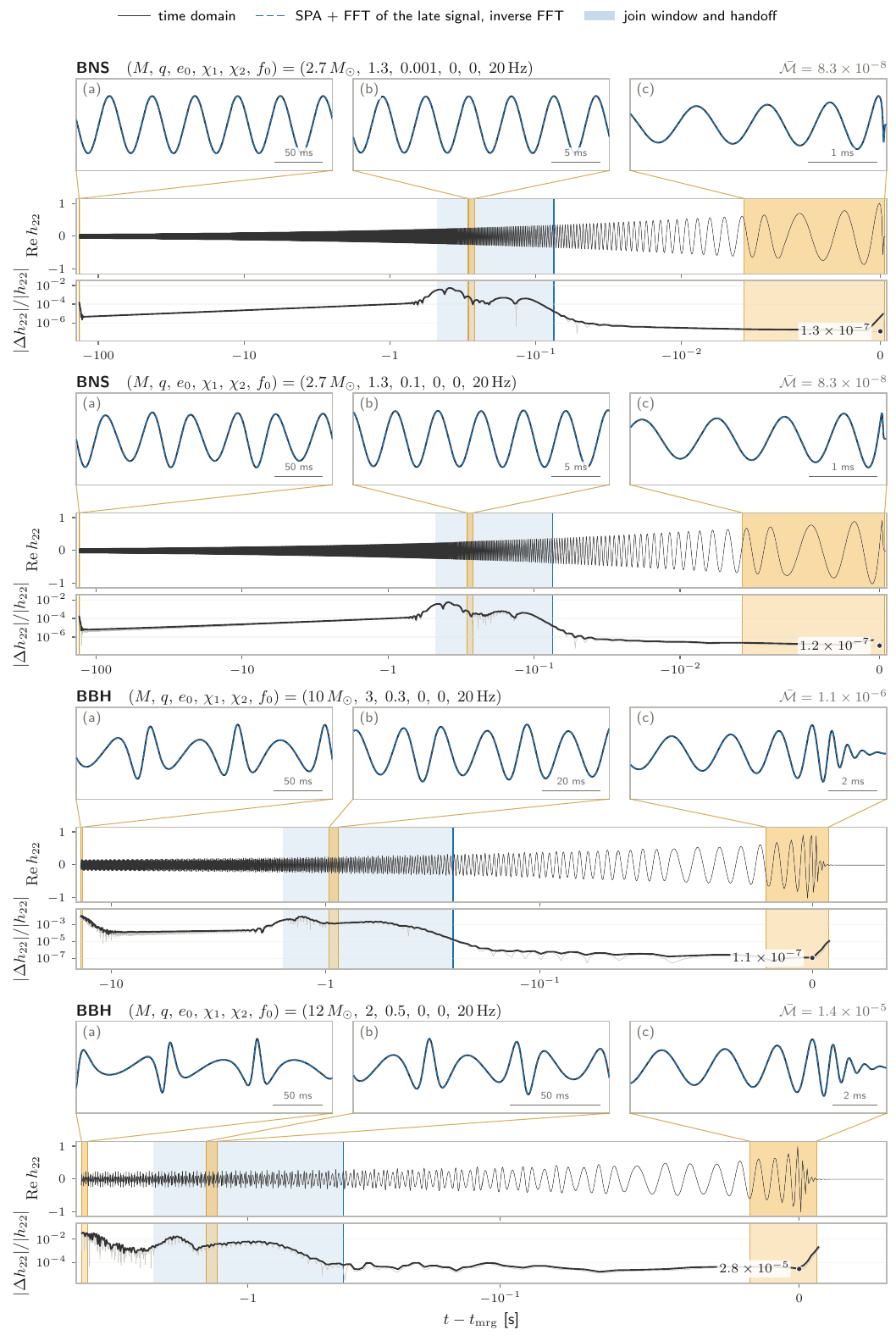}
    \caption{\label{fig:fd_join} The $(2,2)$ mode computed in the \ac{fd} mapped to \ac{td} via inverse
    \ac{fft} (dashed blue) against the \ac{td} mode (solid black) for the four systems of Sec.~\ref{sec:fd_examples}, top to
    bottom. Insets show $\mathrm{Re}\,h_{22}$ early in the
    inspiral, inside the join window and through merger. Below them, the full $\mathrm{Re}\,h_{22}$, with the join window in light
    blue and the handoff as a blue line, and the relative difference $|\Delta h_{22}|/|h_{22}|$: raw (light grey) and its largest
    value per cycle (black).
            }
\end{figure*}

As a proof of concept, I test this construction on the $(2,2)$ mode, computing the amplitudes on the nodes together with
their $O(\epsilon)$ corrections. Note that, in contrast to the full \ac{td} waveform,
the steps outlined in this section are carried out via a simple \texttt{python} implementation once the
output of the model is available.

I consider four nonspinning
systems evolved from $20$~Hz at \ac{nnlo}+ \ac{pa} order:
\begin{itemize}
    \item two \ac{bns}-like binaries ($M = 2.7\,M_\odot$, $q = 1.3$, $\Lambda_{1,2} = 400$) at $e_0 = 10^{-3}$ and $0.1$,
    \item two \acp{bbh} with $(M/M_\odot, q, e_0) = (10, 3, 0.3)$ and $(12, 2, 0.5)$. 
\end{itemize}
I compare the joined \ac{fd} waveform against the full \ac{td} one by taking its inverse \ac{fft} (Fig.~\ref{fig:fd_join}).
The mismatch $\bar{\mathcal{M}}$ is $\FDmmBnsEcoLow$ ($e_0 = 10^{-3}$)
and $\FDmmBnsEcoHigh$ ($e_0 = 0.1$) for the \acp{bns}, and $\FDmmBbhEcoLow$ and
$\FDmmBbhEcoHigh$ for the \acp{bbh} using a flat \ac{psd}.
The relative amplitude difference at merger is at most $\FDresMrgMax$ across all systems, and about $10^{-4}$
during the inspiral, in the \ac{spa} regime.

In terms of computational cost, the \ac{fd} and \ac{td} routes are broadly comparable for the \acp{bns}.
The call to the model is the dominant cost, with the evaluation of the interpolants at every frequency (which is slow in
Python) being a close second. 
The advantage of the \ac{fd} route lies instead in the fact that it can be evaluated on an arbitrary set of
frequencies: on $10^2$--$10^4$ frequencies, as required e.g.\ by relative
binning~\cite{Zackay:2018qdy,Leslie:2021ssu,Cornish:2021lje}, multibanding~\cite{Vinciguerra:2017ngf,Morisaki:2021ngj} or
reduced-order quadratures~\cite{Canizares:2014fya}, the second step can be made significantly
cheaper. For shorter signals, such as the \ac{bbh} at $e_0 = 0.3$, the \ac{fd} route does not pay off, since the
evaluations on the nodes cost more than the reconstruction they replace. 

\section{Conclusions}
\label{sec:conclusion}
In this work I applied the method of variation of constants to the \ac{eob} model \dali{}, obtaining a fast and
accurate description of the inspiral of eccentric compact binaries. The orbital elements are split into a secular part,
evolved on the radiation-reaction timescale, and periodic oscillations that are reconstructed afterwards. Once the binary
is close to plunge, the standard \acp{eom} take over, and merger and ringdown are those of the underlying model.
The construction applies to a production model with precessing spins and tides, makes no assumption on the form of the Hamiltonian
or of the radiation reaction, and is carried to \ac{nnlo}+. Most importantly, it is formulated in terms of equinoctial elements,
for which the circular orbit is a regular point.

The resulting waveforms are in good agreement with those of the full \ac{ode} integration, with an orbital
dephasing at the handoff below $\SwDphiHandoffMaxBns$ rad for $e_0 \leq 0.6$. The gain in speed is largest
for systems with long inspirals, such as \acp{bns} starting from low frequencies.
For the systems that accumulate fewer than $N_{\rm crit}(e_0)$ cycles in band, the short signals of
Sec.~\ref{sec:benchmarks}, the approach does not outperform the standard \ac{ode} integration in terms of computational cost.
Applied to initial data, the same machinery removes the spurious eccentricity that conservative initial
conditions leave in the quasi-circular limit, lowering the floor from $\sim 10^{-3}$ to $\sim 7 \times 10^{-8}$, well below the values
reached by eccentricity-reduced \ac{nr} simulations.
Finally, the method also provides the building blocks for constructing frequency-domain waveforms directly from the secular evolution,
bypassing the need for dense time-domain sampling of the full dynamics, opening the possibility of employing
techniques such as relative binning and reduced-order modeling also for eccentric \ac{eob} waveforms.

A number of limitations remain. For short signals, the secular evolution is now the most expensive part of the \ac{pa} stage,
most of it being spent on second-order corrections that are rebuilt at each run; being smooth functions of the mean elements, these could be
tabulated offline once the model is fixed, as is standard for \acp{emri}~\cite{Chua:2020stf,Katz:2021yft,Chapman-Bird:2025xtd,Lynch:2026ibo}.
For long signals, instead, the larger share goes into the reconstruction and the multipoles (Sec.~\ref{sec:benchmarks}).
Since the modes can be written as sums of harmonics of the mean angles, whose
amplitudes depend only on the mean elements (Eq.~\eqref{eq:fd_modes}), they could instead be
either computed directly as done for the frequency-domain waveforms of Sec.~\ref{sec:fd}, with the full model needed
only close to the handoff, or tabulated offline~\cite{Lynch:2026eyw}.
Along these lines, further optimizations and development
of the \ac{fd} framework itself will be necessary, from reducing the cost
of the $O(\epsilon)$ corrections to extending the construction to all modes and implementing it fully in C.
Finally, the method evolves a planar orbit: precessing spins enter through the spin evolution and the twist
on top of it (Sec.~\ref{sec:dali}), and a fully 3D dynamics
would require adding the orientation of the orbital plane to the slowly varying elements~\cite{Lynch:2024ohd,Klein:2021jtd,Klein:2018ybm}.
This twisting approximation was studied in the context of eccentric orbits in Refs.~\cite{Gamba:2024cvy, Thomas:2026qrg},
but its potential limitations still require further investigation.
Unbound orbits, on the other hand, have no radial period to average over and call for a different treatment altogether~\cite{BernuzziInPrep}.
I leave these developments, and the application of the model to the analysis of \ac{gw} events,
to future work.

Concurrently with this work, Lynch et al.~\cite{Lynch:2026eyw} carried out both of the tabulations discussed above, for the
averaged equations and for the harmonics of the modes, after reformulating the near-identity transformations of
Ref.~\cite{Lynch:2026ibo} in terms of the Cartesian components of the eccentricity vector, which, like the equinoctial elements
used here, are regular in the circular limit. The waveform is then assembled directly on the output time grid at a fixed cost per sample.
Applied to a nonspinning \ac{eob} model implemented in Mathematica, their waveforms reach median mismatches of $\sim 10^{-8}$
at 2\ac{pa} order, with speedups of up to two orders of magnitude with respect to the full \ac{eob} evolution in the same implementation.
The approach of Ref.~\cite{Lynch:2026eyw} and the one presented here are complementary, since an offline tabulation makes each waveform
very cheap once the model is fixed, while computing the averages on the fly applies unchanged to any version of the model.

\section*{Acknowledgments}
R.~G. warmly thanks D.~Chiaramello, S.~Albanesi, S.~Bernuzzi and A.~Nagar for useful discussions and feedback
on this work, as well as all the participants of the
``Binary compact objects on eccentric orbits and their gravitational-wave signatures'' workshop,
held in Corfu at the Corfu Summer Institute.
R.~G. is also grateful to P.~Lynch, A.~Buonanno, A.~Gamboa and M.~van de Meent for their comments on the manuscript.
Finally, R.~G. also thanks C.~R.~Jepsen for inspiration
during the preparation of this work.
R.~G. is supported by NSF Grant PHY-2020275
(Network for Neutrinos, Nuclear Astrophysics, and Symmetries (N3AS)).
\paragraph*{AI statement} AI tools were used by the author to assist in drafting and editing
the manuscript, as well as for the implementation of the algorithms described herein.
Both code and manuscript were carefully reviewed by the author to ensure accuracy and consistency,
and all responsibilities for the content remain with the author.
\paragraph*{Data availability} The code used to generate the results presented in this work will be
made publicly available at \url{https://github.com/teobresums/teobresums}, tagged with the arXiv number of this work.
\paragraph*{Software} This work made use of the following software packages:
\texttt{numpy}~\cite{Harris:2020xlr}, \texttt{scipy}~\cite{Virtanen:2019joe}, \texttt{matplotlib}~\cite{Hunter:2007ouj},
\texttt{PyART}~\cite{PyART}, \texttt{pycbc}~\cite{pycbc}, \texttt{lalsuite}~\cite{lalsuite,Wette:2020air}, \texttt{pyseobnr}~\cite{Mihaylov:2023bkc},
\texttt{gw\_eccentricity}~\cite{Shaikh:2023ypz,Shaikh:2025tae}.

\appendix
\section{The osculating conditions}
\label{app:osc}
In this appendix I review the osculating conditions for both the parameterization of Sec.~\ref{sec:dgi_summary} and the
equinoctial one of Sec.~\ref{sec:equinoct} and show where the two differ. 

\subsection{The condition}
Let $x = (r, p_{r_*}, \varphi, p_\varphi)$. A conservative orbit is labeled by the constants $c_\mu$ and by
two angles $\theta$, an anomaly and an azimuthal angle, through
\begin{equation}
    x = \mathcal{X}(\theta; c_\mu) ,
\end{equation}
such that $\dot c_\mu = 0$ and $\dot\theta = \omega_\theta(c_\mu)$ reproduce Hamilton's equations, i.e.\
$(\partial \mathcal{X}/\partial\theta)\, \omega_\theta = \dot x_{\rm cons}$. With dissipation the same functional form
is kept, with the constants and the angles now functions of time,
\begin{subequations}
    \label{eq:osc_setup}
    \begin{align}
    x(t) &= \mathcal{X}\bigl(\theta(t); c_\mu(t)\bigr) , \label{eq:osc_map} \\
    \dot x &= \dot x_{\rm cons} + \bigl(0, \sqrt{A/B}\, \mathcal{F}_r, 0, \mathcal{F}_\varphi\bigr) , \label{eq:osc_rates}
    \end{align}
\end{subequations}
the latter from Eqs.~\eqref{eq:eom_tortoise}. Differentiating Eq.~\eqref{eq:osc_map} and inserting
Eq.~\eqref{eq:osc_rates}, the conservative parts cancel by construction, and what is left are the osculating conditions,
\begin{equation}
    \label{eq:osc_general}
    \frac{\partial \mathcal{X}}{\partial c_\mu}\, \dot c_\mu + \frac{\partial \mathcal{X}}{\partial \theta}\,
    \bigl(\dot\theta - \omega_\theta\bigr) = \bigl(0, \sqrt{A/B}\, \mathcal{F}_r, 0, \mathcal{F}_\varphi\bigr) .
\end{equation}
In words, the change of the elements, together with any correction to the rates of the angles, must reproduce the forces exactly.
The components along $r$ and $\varphi$ vanish, since the forces do not enter $\dot r$ and $\dot\varphi$: this is
the requirement that the perturbed trajectory be tangent to the unperturbed one. The unknowns to determine are then
$\dot c_\mu$ and the corrections $\dot\theta - \omega_\theta$.

\subsection{$E$ and $p_\varphi$ parameterization}
In Sec.~\ref{sec:dgi_summary} the constants are $c_\mu = (E, p_\varphi)$ and the angles are $(l, \lambda)$. The radial sector reads
$r = S(l; c_\mu)$ and $p_{r_*} = P(l; c_\mu)$, with $P$ the radial momentum along the orbit. Since 
$\dot p_\varphi = \mathcal{F}_\varphi$, the unknowns left are $\dot E$ and
$F_l = \dot l - n$: two unknowns for the two radial components, $\dot{r}$ and $\dot{p}_{r_*}$. The first gives
\begin{equation}
    \begin{aligned}
    &\frac{\partial S}{\partial c_\mu}\, \dot c_\mu + \frac{\partial S}{\partial l}\, F_l = 0 \\
    &\Longrightarrow \quad F_l = -\frac{\partial S}{\partial c_\mu}\, \dot{c}_\mu \left(\frac{\partial S}{\partial l}\right)^{-1} ,
    \end{aligned}
\end{equation}
while the one along $p_{r_*}$ fixes $\dot E$, and is equivalent to the energy balance of Eq.~\eqref{eq:cdot}. The azimuthal component, with
$\varphi = \lambda + V(l; c_\mu)$, gives
\begin{equation}
    F_\lambda = \dot\lambda - (1+k)\, n = -\frac{\partial V}{\partial l}\, F_l - \frac{\partial V}{\partial c_\mu}\, \dot{c}_\mu ,
\end{equation}
and these are Eqs.~\eqref{eq:Fl_general}. No freedom is left, and $F_l$ does not vanish.

\subsection{Equinoctial parameterization}
In Sec.~\ref{sec:equinoct} the constants are $c_\mu = (\alpha, \beta, p_\varphi)$ and the anomaly is $X$. The radial sector reads
$r = R(X; c_\mu)$ and $p_{r_*} = P(X; c_\mu)$, with $R$ and $P$ given by Eqs.~\eqref{eq:eq_param} and~\eqref{eq:prstar_of_X},
and again $\dot p_\varphi = \mathcal{F}_\varphi$. The unknowns are now $\dot\alpha$, $\dot\beta$ and $F_X = \dot X - \omega_X$,
with $\omega_X = (dt/dX)^{-1}$ the conservative rate of Eq.~\eqref{eq:dtdX}, and the two radial components read
\begin{equation}
    \begin{aligned}
    &\frac{\partial R}{\partial c_\mu}\, \dot c_\mu + \frac{\partial R}{\partial X}\, F_X = 0 , \\
    &\frac{\partial P}{\partial c_\mu}\, \dot c_\mu + \frac{\partial P}{\partial X}\, F_X = \sqrt{A/B}\, \mathcal{F}_r \, ,
    \end{aligned}
\end{equation}
that is, two equations for three unknowns.
One condition must therefore be chosen, and it is convenient to set
\begin{equation}
    F_X = 0 ,
\end{equation}
i.e.\ that $r$ and $p_{r_*}$ be unchanged by the variation of the elements at fixed $X$. With this choice, the two equations above
are the system of Eq.~\eqref{eq:eq_osc} for $(\dot\alpha, \dot\beta)$. The azimuthal component, with $\varphi = \lambda + V(X; c_\mu)$,
and the definition $\Lambda = X + T(X; c_\mu)$ of Eq.~\eqref{eq:TV_maps} then give
\begin{equation}
    \label{eq:phase_forcings}
    F_\Lambda = \frac{\partial T}{\partial c_\mu}\, \dot c_\mu , \qquad F_\lambda = -\frac{\partial V}{\partial c_\mu}\, \dot c_\mu .
\end{equation}

Both parameterizations enforce Eq.~\eqref{eq:osc_general}, and differ
mainly in how the periastron evolution is represented. In the standard picture, the periastron direction is not an explicit element;
its motion is absorbed into the anomaly yielding $F_l \neq 0$. In the equinoctial parameterization, conversely,
$F_X = 0$ implies $\dot w \neq 0$. 
This choice ensures the regularity of equinoctial elements in the circular limit.
Enforcing $\dot w = 0$ while allowing $F_X \neq 0$ would reproduce the same construction of Sec.~\ref{sec:dgi_summary}
with $F_X \sim 1/e$, because $\partial R/\partial X \propto r_a$.

\section{Equations at higher orders}
\label{app:higher}
This appendix lists the equations of the \ac{nlo}, \ac{nnlo} and \ac{nnlo}+ levels in the variables of
Sec.~\ref{sec:equinoct}. Subscripts after a comma denote partial derivatives with respect to the elements at fixed $X$, e.g.\
$n_{,\mu} = \partial n/\partial c_\mu$, and a subscript $X$ the derivative with respect to $X$ at fixed elements;
$\omega_\lambda \equiv (1+k)\,n$. Every quantity is evaluated on the mean orbit $\bar c_\mu$ at the anomaly $X_{\rm m}$ of
Eq.~\eqref{eq:Xm}, and $\langle \cdot \rangle$ is the time average over its radial period.

The mean rates of Eq.~\eqref{eq:eq_rates} are
\begin{equation}
    \label{eq:rates_eq}
    \begin{aligned}
    \bar F^{(1)}_\mu &= \langle F_\mu \rangle , \qquad
    \bar F^{(2)}_\mu = \langle \delta F^{(2)}_\mu \rangle , \qquad
    \bar F^{(3)}_\mu = \langle \delta F^{(3)}_\mu \rangle , \\
    \bar F^{(1)}_\Lambda &= \langle F_\Lambda \rangle , \qquad
    \bar F^{(2)}_\Lambda = \Bigl\langle \tfrac12\, n_{,\mu\nu}\, \tilde c^{(1)}_\mu \tilde c^{(1)}_\nu
        + \delta F^{(2)}_\Lambda \Bigr\rangle , \\
    \bar F^{(1)}_\lambda &= \langle F_\lambda \rangle , \qquad
    \bar F^{(2)}_\lambda = \Bigl\langle \tfrac12\, \omega_{\lambda,\mu\nu}\, \tilde c^{(1)}_\mu \tilde c^{(1)}_\nu
        + \delta F^{(2)}_\lambda \Bigr\rangle ,
    \end{aligned}
\end{equation}
and the oscillating parts $\tilde c^{(i)}_\mu$, $\tilde\Lambda^{(i)}$ and $\tilde\lambda^{(i)}$ of the map are the
zero-average primitives of Eq.~\eqref{eq:gen1}, with $F_\mu$ replaced by the sources
\begin{equation}
    \label{eq:sources_eq}
    \begin{aligned}
    S^{(1)}_\mu &= F_\mu , \\
    S^{(1)}_\Lambda &= \delta n^{(1)} + F_\Lambda , \qquad
    S^{(1)}_\lambda = \delta\omega_\lambda^{(1)} + F_\lambda , \\
    S^{(2)}_\mu &= \delta F^{(2)}_\mu - \mathcal{D}^{(1)} \tilde c^{(1)}_\mu , \\
    S^{(2)}_\Lambda &= \delta n^{(2)} + \delta F^{(2)}_\Lambda - \mathcal{D}^{(1)} \tilde\Lambda^{(1)} , \\
    S^{(2)}_\lambda &= \delta\omega_\lambda^{(2)} + \delta F^{(2)}_\lambda - \mathcal{D}^{(1)} \tilde\lambda^{(1)} , \\
    S^{(3)}_\mu &= \delta F^{(3)}_\mu - \mathcal{D}^{(2)} \tilde c^{(1)}_\mu - \mathcal{D}^{(1)} \tilde c^{(2)}_\mu , \\
    S^{(3)}_\Lambda &= \delta n^{(3)} + \delta F^{(3)}_\Lambda
        - \mathcal{D}^{(2)} \tilde\Lambda^{(1)} - \mathcal{D}^{(1)} \tilde\Lambda^{(2)} , \\
    S^{(3)}_\lambda &= \delta\omega_\lambda^{(3)} + \delta F^{(3)}_\lambda
        - \mathcal{D}^{(2)} \tilde\lambda^{(1)} - \mathcal{D}^{(1)} \tilde\lambda^{(2)} ,
    \end{aligned}
\end{equation}
where, for an oscillating part $g$,
\begin{equation}
    \label{eq:D_eq}
    \begin{aligned}
    \mathcal{D}^{(k)} g &= \frac{\partial g}{\partial \bar c_\nu}\bigg|_{\bar\Lambda} \bar F^{(k)}_\nu
        + \frac{\partial g}{\partial \bar\Lambda}\, \bar F^{(k)}_\Lambda , \\
    \frac{\partial g}{\partial \bar\Lambda} &= \frac{g_X}{n\, dt/dX} , \qquad
    \frac{\partial g}{\partial \bar c_\nu}\bigg|_{\bar\Lambda} = g_{,\nu} - \frac{g_X\, T_{,\nu}}{n\, dt/dX} .
    \end{aligned}
\end{equation}
The variations of a forcing $f \in \{F_\mu, F_\Lambda, F_\lambda\}$ are
\begin{equation}
    \label{eq:df_eq}
    \begin{aligned}
    \delta f^{(2)} &= f_{,\nu}\, \tilde c^{(1)}_\nu + f_X\, \delta X^{(1)} , \\
    \delta f^{(3)} &= f_{,\nu}\, \tilde c^{(2)}_\nu + f_X\, \delta X^{(2)}
        + \tfrac12\, f_{,\nu\sigma}\, \tilde c^{(1)}_\nu \tilde c^{(1)}_\sigma \\
        &\quad + f_{,\nu X}\, \tilde c^{(1)}_\nu\, \delta X^{(1)} + \tfrac12\, f_{XX} \bigl(\delta X^{(1)}\bigr)^2 ,
    \end{aligned}
\end{equation}
those of a frequency $g \in \{n, \omega_\lambda\}$ are
\begin{equation}
    \label{eq:dg_eq}
    \begin{aligned}
    \delta g^{(1)} &= g_{,\mu}\, \tilde c^{(1)}_\mu , \qquad
    \delta g^{(2)} = g_{,\mu}\, \tilde c^{(2)}_\mu + \tfrac12\, g_{,\mu\nu}\, \tilde c^{(1)}_\mu \tilde c^{(1)}_\nu , \\
    \delta g^{(3)} &= g_{,\mu}\, \tilde c^{(3)}_\mu + g_{,\mu\nu}\, \tilde c^{(1)}_\mu \tilde c^{(2)}_\nu
        + \tfrac16\, g_{,\mu\nu\sigma}\, \tilde c^{(1)}_\mu \tilde c^{(1)}_\nu \tilde c^{(1)}_\sigma ,
    \end{aligned}
\end{equation}
and the displacements of the anomaly are
\begin{equation}
    \label{eq:dX_eq}
    \begin{aligned}
    n \frac{dt}{dX}\, \delta X^{(1)} &= \tilde\Lambda^{(1)} - T_{,\mu}\, \tilde c^{(1)}_\mu , \\
    n \frac{dt}{dX}\, \delta X^{(2)} &= \tilde\Lambda^{(2)} - T_{,\mu}\, \tilde c^{(2)}_\mu
        - \tfrac12\, T_{XX} \bigl(\delta X^{(1)}\bigr)^2 \\
        &\quad - T_{,\mu X}\, \tilde c^{(1)}_\mu\, \delta X^{(1)}
        - \tfrac12\, T_{,\mu\nu}\, \tilde c^{(1)}_\mu \tilde c^{(1)}_\nu .
    \end{aligned}
\end{equation}
The map from osculating to mean elements is solved by fixed-point iteration, with $X_{\rm m}$ updated at each iteration,
\begin{equation}
    \label{eq:inv_eq}
    \begin{aligned}
    \bar c_\mu &= c_\mu - \sum_i \epsilon^i\, \tilde c^{(i)}_\mu(\bar c_\sigma, \bar\Lambda) , \\
    \bar\Lambda &= \Lambda - \sum_i \epsilon^i\, \tilde\Lambda^{(i)} , \qquad
    \bar\lambda = \lambda - \sum_i \epsilon^i\, \tilde\lambda^{(i)} .
    \end{aligned}
\end{equation}
\Ac{lo} uses $\bar F^{(1)}_\mu$ only, with the angles advancing at $n$ and $\omega_\lambda$ and no map. \Ac{nlo} adds
$\bar F^{(2)}_\mu$, $\bar F^{(1)}_\Lambda$, $\bar F^{(1)}_\lambda$ and the first-order map; \ac{nnlo} adds $\bar F^{(3)}_\mu$,
$\bar F^{(2)}_\Lambda$, $\bar F^{(2)}_\lambda$ and the second-order map; \ac{nnlo}+ adds the third-order map.

\section{Implementation details}
\label{app:techimp}
\begin{figure*}
    \centering
    \includegraphics[width=0.8\textwidth]{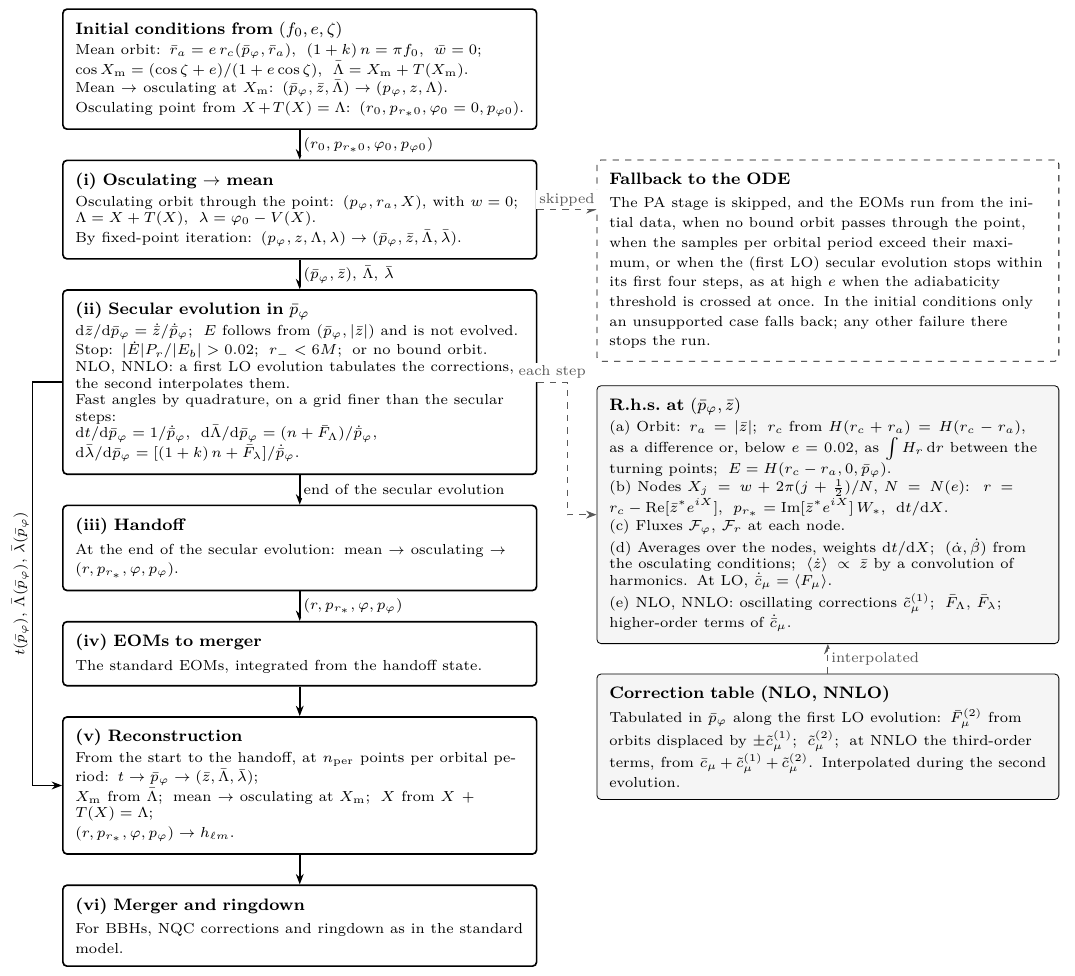}
    \caption{Schematic representation of the algorithm. Steps (i)--(vi) are those of the workflow of Sec.~\ref{sec:equinoct}.}
    \label{fig:schematic}
\end{figure*}

This appendix describes the numerical implementation of the \ac{pa} stage, shown schematically in Fig.~\ref{fig:schematic}.

\subsection{The conservative orbit}
Given $(r_a, p_\varphi)$, the orbit follows from energy conservation. The radius $r_c$ is the root of
\begin{equation}
    \label{eq:energy_conservation}
    \delta H(r_c) \equiv H(r_c + r_a, 0, p_\varphi) - H(r_c - r_a, 0, p_\varphi) = 0 ,
\end{equation}
which equates the energies of the two turning points, and the energy is $E = H(r_c - r_a, 0, p_\varphi)$. At each anomaly, the
radial momentum of Eq.~\eqref{eq:prstar_of_X} follows from $H(r, p_{r_*}, p_\varphi) = E$, which gives
\begin{equation}
    \label{eq:gap}
    W_*^2 = \frac{E - H(r, 0, p_\varphi)}{(dr/dX)^2\, \Phi} , \qquad
    \Phi = \frac{1}{p_{r_*}^2} \int_0^{p_{r_*}^2} \partial_s H(r, s, p_\varphi)\, ds ,
\end{equation}
where $H(r, s, p_\varphi)$ denotes the Hamiltonian as a function of the squared radial momentum $s$.
Since $\Phi$ depends on $p_{r_*} = (dr/dX)\, W_*$, Eq.~\eqref{eq:gap} is
solved for $W_*$ by fixed-point iteration, starting from $\Phi = \partial H/\partial (p_{r_*}^2)$ at $p_{r_*} = 0$; a few
iterations typically suffice. The time along the orbit then follows from
\begin{equation}
    \label{eq:dtdX}
    \frac{dt}{dX} = \frac{dr/dX}{\dot r} = \frac{1}{2 \sqrt{A/B}\; W_*\; \partial H/\partial(p_{r_*}^2)} ,
\end{equation}
which replaces Eq.~\eqref{eq:dtdchi} and is finite at the turning points and at $e = 0$.

Both $\delta H$ and the numerator of Eq.~\eqref{eq:gap} are differences of the Hamiltonian, which become noisy to compute
as $r_a \to 0$. For $e < 0.02$ each is computed instead as the integral of $H_r$:
between the two turning points for $\delta H$, with an eight-point Gauss-Legendre rule, and along the
orbit from periastron for $E - H(r, 0, p_\varphi)$. The function $\Phi$ applies the same device in $p_{r_*}^2$.

The osculating system of Eq.~\eqref{eq:eq_osc} requires the derivatives of $r_c$ and $W_*$ with respect to $p_\varphi$ and $r_a$.
Those of $r_c$ follow from the implicit differentiation of Eq.~\eqref{eq:energy_conservation}, as ratios of combinations of
$H_{p_\varphi}$ and $H_r$ at the two turning points; since numerator and denominator vanish together as $e \to 0$, their circular
limits are used at very small eccentricity. Those of $W_*$, taken at fixed $r$, follow from differentiating Eq.~\eqref{eq:gap}, and
depend on the derivatives of $r_c$ through those of $E$.

\subsection{Orbit-averaged quantities}
Orbit averages of a function $g(r, p_{r_*}, p_\varphi)$, the time averages over a radial period of Eq.~\eqref{eq:avg_chi},
are evaluated on $N$ nodes in the anomaly,
\begin{equation}
    \label{eq:avg_nodes}
    \begin{aligned}
    \langle g \rangle &= \frac{1}{P_r} \int_0^{2\pi} g\, \frac{dt}{dX}\, dX
                       \simeq \frac{2\pi}{N P_r} \sum_{j=0}^{N-1} g(X_j) \frac{dt}{dX}\Big|_{X_j} , \\
    X_j &= w + \frac{2\pi}{N}\bigl(j + \tfrac12\bigr) .
    \end{aligned}
\end{equation}
The nodes are offset by half a step from periastron and their number $N$ is taken as a power of two, hence even, so that none falls on a
turning point, where the ratio in Eq.~\eqref{eq:gap} and the osculating
system of Eq.~\eqref{eq:eq_osc} take the form $0/0$. Since the integrand is periodic, the error of this rule comes only from its
harmonics at multiples of $N$, which take the same value at every node and are therefore indistinguishable from a constant (``aliasing''), 
and it decreases exponentially with $N$~\cite{Trefethen:2014}. Moreover, 
since $r$ and $p_{r_*}$ depend on the anomaly only through $r_a \cos X$ and $r_a \sin X$, 
the $m$-th harmonic of the integrand first appears at order $r_a^m$\footnote{The analogue of the d'Alembert characteristic of celestial mechanics~\cite{Brouwer:1961,Murray:1999},
see also Refs.~\cite{Peters:1963ux,Yunes:2009yz}}, which is $O(e^m)$.
Requiring the first of these harmonics, of relative size $\mathcal{C} e^N$, to be below a
tolerance gives
\begin{equation}
    N \ge \frac{\ln(\mathcal{C}/\mathrm{tol})}{\ln(1/e)} , \qquad \mathrm{tol} = 10^{-11}, \quad \mathcal{C} = 30 ,
\end{equation}
where the constant $\mathcal{C}$ is fitted to the error measured on the orbit-averaged energy flux.

The only exception to the procedure above is the average $\langle \dot z \rangle$.
To maintain accuracy at small $e$, it is convenient to compute it as a convolution of harmonics.
Writing $\dot z = C(X)\, e^{iX}$,
\begin{equation}
    \label{eq:zdot_conv}
    \langle \dot z \rangle = n\, e^{i w} \sum_m \hat C_m\, \hat\tau_{-(1+m)} , \qquad
    \hat\tau_m = \kappa_{|m|}\, r_a^{|m|} ,
\end{equation}
where $\hat C_m$ and $\hat\tau_m$ are the harmonics of $C$ and of $dt/dX$ in $\chi$, and the coefficients $\kappa_m$
are finite as $e \to 0$. For the same reason as before $\hat{C}_m = O(r_a^{|m|})$, and
each term entering the sum vanishes identically at $z = r_a = 0$.

\subsection{The \ac{nit} and the higher-order rates}
The oscillating parts $\tilde c^{(i)}_\mu$ are the zero-average periodic primitives of their sources, computed on the nodes.
Rather than evolving them, I compute them from the $F_\mu$ themselves.
At first order, the oscillating part is given by
\begin{equation}
    \label{eq:gen1}
    \tilde c^{(1)}_\mu(X) = \int^{X} \bigl[ F_\mu - \langle F_\mu \rangle \bigr] \frac{dt}{dX'}\, dX' ,
    \qquad \bigl\langle \tilde c^{(1)}_\mu \bigr\rangle = 0 .
\end{equation}
The integrand is fitted with a Fourier series in $X$, and each harmonic $m \neq 0$ is divided by $i m$.
The constant of integration is then chosen so that the resulting primitive has zero time average.
The oscillating parts of $\Lambda$ and $\lambda$ are obtained in the same way.

The higher-order rates contain derivatives of the forcing with respect to the elements, e.g.\ the term
$(\partial F_\mu/\partial c_\nu)\, \tilde c^{(1)}_\nu$ of $\bar F^{(2)}_\mu$ (App.~\ref{app:higher}),
which arise from Taylor-expanding $F_\mu$ about the mean elements. In practice,
rather than computing the derivatives via finite differences, it is more efficient to
directly ``undo'' the Taylor expansion. This means evaluating the $F_\mu$
on the displaced orbits $\bar c_\mu \pm \tilde c^{(1)}_\mu$, rather than on the mean elements alone.
The terms of the desired order are then isolated by parity and by
subtracting those of lower order: half the difference of the two gives the part linear in $\tilde c^{(1)}_\mu$, which enters
$\bar F^{(2)}_\mu$, and half their sum, minus the forcing on the mean orbit, the quadratic one. The quadratic part enters the second-order rates of the angles, through the frequencies $n$ and
$\omega_\lambda$, and, together with the forcing on the orbit $\bar c_\mu + \tilde c^{(1)}_\mu + \tilde c^{(2)}_\mu$, the third-order rates.

Even though this trick saves a significant amount of time by reducing the number of
flux calls needed, it is still computationally expensive if done at every step of the integrator.
To avoid this burden, at \ac{nlo}, \ac{nnlo} and \ac{nnlo}+ the secular evolution is actually
run \textit{twice} (right column of Fig.~\ref{fig:schematic}):
\begin{itemize}
    \item First, a cheap \ac{lo} evolution is performed to place 12 to 32 ``checkpoints'' along the inspiral
    (spaced every $16$ secular samples).
    \item At each checkpoint, higher-order corrections are calculated and stored as smooth functions of
    $\bar p_\varphi$ and expressed as ratios: the energy and $p_\varphi$ rate corrections normalized by their first-order values,
    $\dot z$ relative to $z$, and the angle rates relative to $n$ and $(1+k)n$.
    \item The derivatives at fixed $\bar\Lambda$ in the $\mathcal{D}^{(k)}$ of Eq.~\eqref{eq:D_eq} are taken by 
    building the oscillating parts at two mean states a small step apart along $\bar F^{(k)}_\nu$, or at two neighboring
    checkpoints for the second-order ones, and differencing them at fixed $\bar\Lambda$.
    \item Finally, the \ac{nlo}, \ac{nnlo} or \ac{nnlo}+ evolution is run from the start,
    with the first-order rates computed at every step and the corrections interpolated
    in $\bar p_\varphi$ with a cubic interpolation.
\end{itemize}

\subsection{Stopping conditions and handoff}
The \ac{pa} stage stops when the adiabaticity parameter, with $E_b$ the binding energy,
\begin{equation}
    \label{eq:adiab}
    \frac{|\dot E|\, P_r}{|E_b|} ,
\end{equation}
exceeds $0.02$, when the periastron $r_-$ falls below $6 M$, or when no bound orbit exists.
The \ac{ode} then takes over from the
state reconstructed at the end of the secular evolution. The \ac{pa} stage is skipped, and the \acp{ode} start from the initial data, 
if Eq.~\eqref{eq:nper} exceeds the chosen maximum or if the secular evolution stops within its first four
steps. The last case occurs at high eccentricity, where the adiabaticity threshold is
crossed almost at once, and accounts for
all such cases in the sweeps of Sec.~\ref{sec:benchmarks}.

\subsection{Reconstruction}
The reconstruction fills the dynamics at a set of sample times, on which the multipoles are then computed and later
interpolated to the output grid, as for the \ac{ode}. At each sample time $t$ it proceeds as follows:
\begin{enumerate}
    \item[(a)] $\bar p_\varphi$ is found numerically on the cubic Hermite interpolant of $t(\bar p_\varphi)$. $\bar\Lambda$
    and $\bar\lambda$ are interpolated in the same way, on the finer grid of their quadratures.
    \item[(b)] $\bar z$ and the harmonics of $T$, $V$ and of the oscillating parts are interpolated in $\bar p_\varphi$, the
    harmonics being measured from the mean periastron, where they vary smoothly.
    \item[(c)] The anomaly $X_{\rm m}$ on the mean orbit is found from Eq.~\eqref{eq:Xm} numerically, and all
    oscillating parts are evaluated there.
    The anomaly $X$ on the osculating orbit differs from $X_{\rm m}$ at $O(\epsilon)$, and using it would leave an error of the
    size of $\tilde c^{(2)}_\mu$.
    \item[(d)] The map from mean to osculating elements, Eq.~\eqref{eq:ic_map}, with the third-order terms at \ac{nnlo}+, gives the osculating elements and
    orbit. $X$ and $\varphi$ follow from Eqs.~\eqref{eq:TV_maps} on the osculating orbit, and $r$, $p_{r_*}$ from
    Eqs.~\eqref{eq:eq_param} and~\eqref{eq:prstar_of_X}.
\end{enumerate}
The samples are spaced by a fixed fraction of the orbital period, with $n_{\rm per}$ samples per period and
\begin{equation}
    \label{eq:nper}
    n_{\rm per} \geq \frac{\ln 10^{7}}{\ln(1/e)} ,
\end{equation}
so that the final interpolation resolves the modulation of the modes over the orbit, whose
$m$-th harmonic is $O(e^m)$; if this exceeds a maximum set by the user, the \ac{pa} stage is skipped.

\section{The circular limit and the quasi-circular \ac{pa} approximation}
\label{app:circular}
In this appendix I show that, for $\bar z = 0$, the equinoctial construction of Sec.~\ref{sec:equinoct} reproduces the
quasi-circular \ac{pa} approximation of Refs.~\cite{Damour:2012ky,Nagar:2018gnk}, at least up to the 1\ac{pa} radial momentum.

Let us start by recalling the basics of the quasi-circular \ac{pa} approximation, in terms of the \acp{eom} of Eq.~\eqref{eq:eom_tortoise}.
The forces are treated as small, and the momenta are expanded in powers of $\epsilon$ at fixed radius. At zeroth order the orbit is
circular and does not inspiral:
\begin{equation}
    \label{eq:j0}
    p_{r_*} = 0 , \qquad p_\varphi = j_0(r) , \qquad \frac{\partial H}{\partial r}\bigl(r, 0, j_0(r)\bigr) = 0 .
\end{equation}
At first order, one asks that $p_\varphi$ keeps following its circular value along the inspiral. Differentiating the second
of Eqs.~\eqref{eq:j0} with respect to time, and using the last of Eqs.~\eqref{eq:eom_tortoise}, fixes $\dot{r}$:
\begin{equation}
    \label{eq:rdot_pa}
    \frac{dj_0}{dr}\, \dot r = \mathcal{F}_\varphi \qquad \Longrightarrow \qquad \dot r = \mathcal{F}_\varphi \Bigl( \frac{dj_0}{dr} \Bigr)^{-1} .
\end{equation}
Since $H$ depends on the radial momentum only through its square,
\begin{equation}
    \frac{\partial H}{\partial p_{r_*}} = 2\, p_{r_*}\, \frac{\partial H}{\partial (p_{r_*}^2)} ,
\end{equation}
where, for small momenta, the last derivative can be evaluated at $p_{r_*} = 0$. With this,
the first of Eqs.~\eqref{eq:eom_tortoise} turns Eq.~\eqref{eq:rdot_pa} into an algebraic equation for $p_{r_*}$ itself:
\begin{equation}
    \label{eq:pa1_prstar}
    p_{r_*}^{\rm 1PA} = \frac{\mathcal{F}_\varphi}{dj_0/dr}\; \frac{1}{2 \sqrt{A/B}\; \partial H/\partial (p_{r_*}^2)} .
\end{equation}
Time and orbital phase then follow by quadratures over the radial grid,
\begin{equation}
    \label{eq:pa_quad}
    t = \int \frac{dr}{\dot r} , \qquad \varphi = \int \Omega\, \frac{dr}{\dot r} .
\end{equation}
Higher orders alternate: even orders correct $p_\varphi$, odd orders correct $p_{r_*}$.

Let us now go back to the equinoctial variables, and consider a binary with $\bar z = 0$. 
The first thing to note is that $\bar z$ then remains zero if at any point it becomes zero, which
means that if an orbit circularizes at any point, it will remain circular.
At first order this can be seen directly from
the system of Eq.~\eqref{eq:eq_osc}. When $z = 0$ the coefficients of Eq.~\eqref{eq:eq_osc_coeffs} lose the terms that
multiply $\alpha$ and $\beta$, and $R_2$ loses the term proportional to $dr/dX$, which vanishes. 
In complex form the solution reads:
\begin{equation}
    \label{eq:zdot_circ}
    \begin{aligned}
    \dot z &= C\, e^{iX} , \qquad C = R_1 - i R_2 , \\
    R_1 &= \frac{\partial r_c}{\partial p_\varphi}\, \mathcal{F}_\varphi , \qquad R_2 = \sqrt{\frac{A}{B}}\, \frac{\mathcal{F}_r}{W_*} ,
    \end{aligned}
\end{equation}
where $C$ does not depend on $X$. Note that $\dot z$ does \textit{not} vanish at $z = 0$:
the osculating elements keep oscillating, however close to circular the orbit is. Its average over a period of $X$,
however, does vanish:
\begin{equation}
    \label{eq:G1_zero}
    \bar F^{(1)}_z(0, \bar p_\varphi) = \bigl\langle C\, e^{iX} \bigr\rangle = 0 \, ,
\end{equation}
so that at \ac{lo} a binary with $\bar z = 0$ remains circular.

Due to symmetry, the same remains true at every order. Indeed, $X$ and $z$ enter $r$ and $p_{r_*}$ only through the combination $z^* e^{iX}$,
meaning that a shift
\begin{equation}
    \label{eq:shift}
    X \to X + \delta , \qquad z \to z\, e^{i\delta} ,
\end{equation}
with $\delta$ a constant does not change the system.
The secular rate of $\bar z$
\begin{equation}
    \label{eq:G_def}
    \dot{\bar z} = \bar F_z\bigl(\bar z, \bar p_\varphi\bigr) , \qquad \bar F_z \equiv \sum_i \epsilon^i\, \bar F^{(i)}_z ,
\end{equation}
must then be given by the same function for $\bar z$ and for $\bar z\, e^{i\delta}$. Since the rate of the latter is the rate
of the former multiplied by $e^{i\delta}$, this means
\begin{equation}
    \label{eq:covariance}
    \bar F_z\bigl(\bar z\, e^{i\delta}, \bar p_\varphi\bigr) = e^{i\delta}\, \bar F_z\bigl(\bar z, \bar p_\varphi\bigr) \qquad \text{for every } \delta ,
\end{equation}
and at $\bar z = 0$, where the left-hand side does not depend on $\delta$, this equality can only hold if
\begin{equation}
    \bar F_z(0, \bar p_\varphi) = 0 .
\end{equation}
The first of Eqs.~\eqref{eq:eq_system} is therefore solved by $\bar z = 0$ at every order, and only the quadratures of its last
two lines are left. 

Although this discussion may appear trivial, it formally demonstrates that since $\bar z = 0$ at all times,
the evolution reduces to a one-parameter family of states labeled by $\bar p_\varphi$, as the quasi-circular \ac{pa}
is labeled by $r$.
Note that this argument assumes that $H$ and the dissipative forces depend only on the instantaneous state $(r, p_{r_*}, p_\varphi)$, and not on any additional degrees of freedom.
This would no longer be the case if, for instance, full dynamical tidal interactions were included: 
a mode that evolves independently carries its own phase, singling out a preferred direction in the orbital plane.
Similarly, transient resonances of \acp{emri}~\cite{Flanagan:2010cd,Lynch:2024ohd}, in-plane spins entering the orbital dynamics, or the presence of a third body would also break the assumption.

Moving on, at $\bar z = 0$ the two turning points coincide and Eq.~\eqref{eq:energy_conservation} is equivalent to
the condition that the radial derivative of the Hamiltonian vanish, which is the last of Eqs.~\eqref{eq:j0}: $r_c$ is the radius
of the circular orbit of angular momentum $\bar p_\varphi$,
\begin{equation}
    r_c = r_{\rm circ}(\bar p_\varphi) \qquad \Longleftrightarrow \qquad \bar p_\varphi = j_0(r_c) .
\end{equation}
Since nothing changes along this orbit, every orbit average is simply the value of its integrand: $\dot{\bar p}_\varphi$ is
$\mathcal{F}_\varphi$ evaluated at $(r_c, 0, \bar p_\varphi)$, and $(1+k)\, n$ is $\Omega$ evaluated at the same point. At \ac{lo}, the quadratures
of Eqs.~\eqref{eq:eq_system} for $t$ and $\bar\lambda$ read
\begin{equation}
    \frac{dt}{d\bar p_\varphi} = \frac{1}{\mathcal{F}_\varphi} , \qquad \frac{d\bar\lambda}{d\bar p_\varphi} = \frac{\Omega}{\mathcal{F}_\varphi} .
\end{equation}
Changing variable from $p_\varphi$ to $r$, and using Eq.~\eqref{eq:rdot_pa}, one finds
\begin{equation}
    dt = \frac{dj_0}{dr} \frac{dr}{\mathcal{F}_\varphi} = \frac{dr}{\dot r} , \qquad d\bar\lambda = \Omega\, \frac{dr}{\dot r} ,
\end{equation}
which are the quadratures of Eq.~\eqref{eq:pa_quad}. The \ac{lo} evolution is therefore the \ac{pa} one,
with $j_0(r)$ used as the coordinate along the inspiral in place of the radius.

At the next order, one has to account for the oscillations about the mean orbit.
Since $\bar z = 0$, the osculating $z$ will be purely oscillatory, while $\mathcal{F}_\varphi$ is constant and coincides with its average. Therefore:
\begin{equation}
    \tilde c^{(1)}_z = \int^X C\, e^{iX'}\, \frac{dX'}{n} = -\frac{i\, C}{n}\, e^{iX} , \qquad \tilde c^{(1)}_{p_\varphi} = 0 .
\end{equation}
Inserting $\tilde c^{(1)}_z$ into Eqs.~\eqref{eq:eq_param} and~\eqref{eq:prstar_of_X}, $r$ and its derivative become
\begin{equation}
    \label{eq:r_circ_recon}
    r = r_c - \frac{{\rm Im}\, C}{n} = r_c + \frac{R_2}{n} , \qquad
    \frac{dr}{dX} = \frac{{\rm Re}\, C}{n} = \frac{R_1}{n} ,
\end{equation}
neither of which depends on $X$.
The radial momentum is 
\begin{equation}
    p_{r_*} = \frac{dr}{dX} W_* = \frac{R_1}{n} W_* \, .
\end{equation}
Using that $\partial r_c/\partial p_\varphi$ in the quasi-circular limit is $(dj_0/dr)^{-1}$, and that Eq.~\eqref{eq:dtdX} at
$z = 0$, where $dt/dX = 1/n$, gives $n = 2\sqrt{A/B}\, W_*\, \partial H/\partial (p_{r_*}^2)$, one finally obtains
\begin{equation}
    p_{r_*} = \frac{\mathcal{F}_\varphi}{dj_0/dr}\; \frac{1}{2 \sqrt{A/B}\; \partial H/\partial (p_{r_*}^2)} = p_{r_*}^{\rm 1PA} ,
\end{equation}
which is exactly the 1PA radial momentum of Eq.~\eqref{eq:pa1_prstar}.

Eq.~\eqref{eq:r_circ_recon} also contains a term which has no counterpart in Eq.~\eqref{eq:pa1_prstar}, namely the offset $r - r_c$.
This quantity is directly proportional to the radial force $\mathcal{F}_r$, and represents the displacement of the orbit due
to it. In the quasi-circular limit and at the order considered here, this term is zero (since $\mathcal{F}_r$ is proportional to
$p_{r_*}$, which vanishes on the mean orbit), but it would contribute at higher orders, where $p_{r_*} \neq 0$.

In conclusion, the \ac{lo} equinoctial evolution reproduces the time and phase of the quasi-circular \ac{pa} approximation, and the first-order map, i.e.\ \ac{nlo}, adds its 1\ac{pa} radial momentum.
While the expansion to higher orders has not been carried out explicitly, the agreement should persist since both
constructions are fundamentally expansions of the same dynamics in the same small parameter $\epsilon$.

\section{Parameter space investigation for precessing binaries}
\label{app:precessing}
In this appendix I explore the same parameter spaces considered in Sec.~\ref{sec:benchmarks}, also
varying the spin orientations of the binary components. The spin magnitudes are those of the aligned-spin draws,
up to $|\chi_i| \leq 0.9$ for \acp{bbh} and $|\chi_i| \leq 0.1$ for \acp{bns}, while their directions are
sampled uniformly over the unit sphere. The co-precessing $(2,2)$ mode is twisted into all the inertial $\ell = 2$ modes.

\begin{figure*}[tp]
    \centering
    \includegraphics[width=\textwidth,height=0.43\textheight,keepaspectratio]{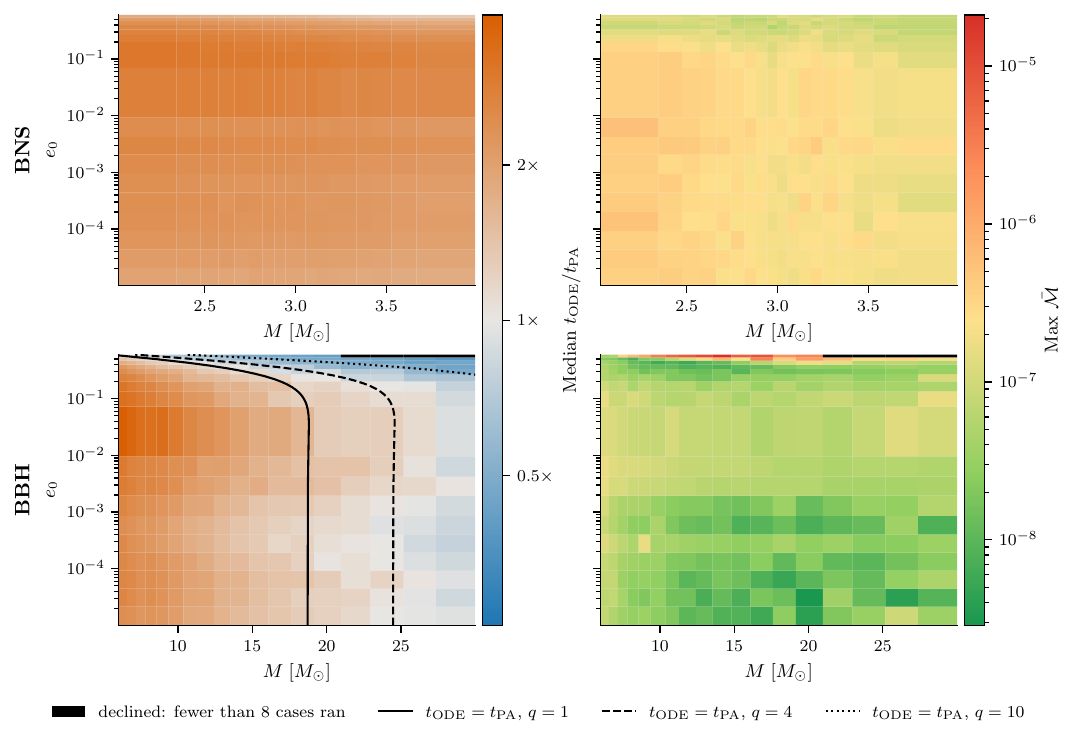}
    \includegraphics[width=\textwidth,height=0.43\textheight,keepaspectratio]{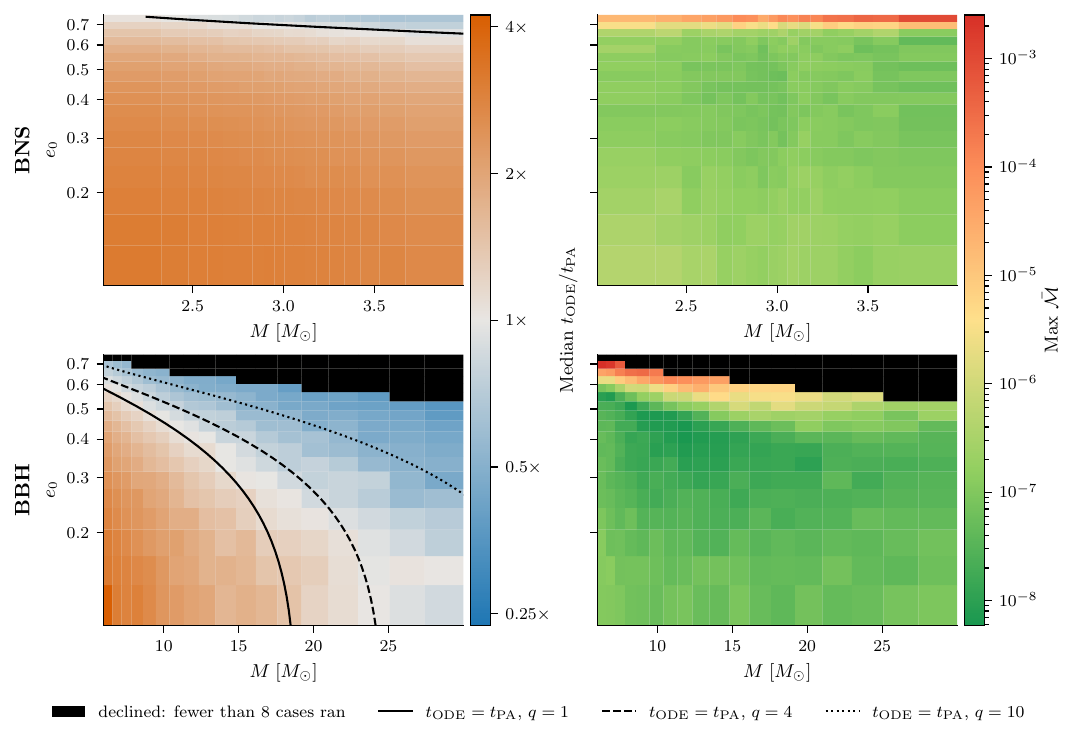}
    \caption{\label{fig:prec_sweep} Same as Figs.~\ref{fig:sweep} and~\ref{fig:sweep_highecc}, for the precessing binaries of
    App.~\ref{app:precessing}: low-eccentricity (top) and high-eccentricity (bottom) parameter spaces. In each, median speedup (left)
    and largest mismatch (right), the latter not maximized over time and phase and computed at an inclination of $\pi/3$. The black
    lines are the $t_{\rm ODE} = t_{\rm PA}$ boundaries of the aligned-spin parameter spaces.}
\end{figure*}

Results are presented in Fig.~\ref{fig:prec_sweep}, which follows the same
format as those in Sec.~\ref{sec:benchmarks}. The only difference, beyond the inclusion of precession,
is that -- for computational reasons -- I compute overlaps rather than mismatches. This means that
instead of Eq.~\eqref{eq:mismatch}, I skip maximization over time and phase, and compare
$h = h_+ - i h_\times$ directly with a flat spectrum, at an inclination of $\pi/3$ (the values at $0$ and $\pi/4$
differ by a few percent).
I also do not maximize over in-plane spin rotations~\cite{Pratten:2020ceb}, differently from what would be done in a full precessing analysis.
The values reported are therefore conservative estimates.

Despite the above caveats, for $e_0 \leq 0.6$ mismatches are at most $\SwPrecMmMaxBbhLow$ for \acp{bbh}
and $\SwPrecMmMaxBnsLow$ for \acp{bns}, with medians of $\SwPrecMmMedBbhLow$ and $\SwPrecMmMedBnsLow$.
These numbers are in good agreement with those obtained from the aligned-spin analysis, indicating that the
spins evolution is not significantly affected by the \ac{pa} evolution, so long as it remains faithful to the underlying \ac{ode} one.
Above $e_0 = 0.6$, $\SwPrecNBigBbh$ of the $\SwPrecNHighBbh$ \acp{bbh} and $\SwPrecNBigBns$ of the $\SwPrecNHighBns$ \acp{bns}
have mismatches above $10^{-4}$, up to $\SwPrecMmMaxBbhHigh$ and $\SwPrecMmMaxBnsHigh$, all with $e_0 > \SwPrecBigEminBbh$
and $e_0 > \SwPrecBigEminBns$, respectively. These are not due to precession: for the binaries with the largest values,
removing the in-plane spin components leaves the mismatch unchanged. They reflect the dephasing between the \ac{pa}
and the \ac{ode} evolutions at the highest eccentricities, which the maximization over time and phase of
Eq.~\eqref{eq:mismatch} largely absorbs in Sec.~\ref{sec:benchmarks}.
In terms of efficiency, once more, I obtain similar figures as Sec.~\ref{sec:benchmarks}, with the \ac{pa} technique delivering median speedups
of $\times \SwPrecSpeedupBbhLow$ for \acp{bbh} and $\times \SwPrecSpeedupBnsLow$ for \acp{bns} for $e_0 \leq 0.6$, and of
$\times \SwPrecSpeedupBbhHigh$ and $\times \SwPrecSpeedupBnsHigh$ in the high-eccentricity parameter space.

\clearpage  
\input{eccpa.bbl}\end{document}

%% file: symbols.tex

\newcommand{\RG}[1]{{\textcolor{dodgerblue}{{RG: #1}} }}
\newcommand{\JL}[1]{{\textcolor{purple}{{JL: #1}} }}
\newcommand{\DC}[1]{{\textcolor{orange}{{DC: #1}} }}
\newcommand{\NC}[1]{{\textcolor{magenta}{{NC: #1}} }}
\newcommand{\kc}[1]{{\textcolor{purple}{{KC: #1}} }}

\newcommand{\todo}[1]{\textcolor{red}{\texttt{TODO: #1}}} 
\newcommand{\red}[1]{\textcolor{red}{#1}} 
\newcommand{\bl}[1]{\textcolor{blue}{#1}} 
\newcommand{\cor}[2]{\sout{#1}\textcolor{red}{#2}} 
\newcommand{\old}[1]{\textcolor{gray}{\sout{#1}}}
\newcommand{\new}[1]{\red{#1}}

\newcommand{\eo}[0]{\hat{E}_0}
\newcommand{\lo}[0]{\hat{L}_0}
\newcommand{\bphi}[0]{\bar{\varphi}}
\newcommand{\dbphi}[0]{\dot{\bar{\varphi}}}
\newcommand{\dphi}[0]{\dot{\varphi}}
\newcommand{\amrg}[1]{{A_{#1}^{\rm mrg}}}
\newcommand{\anqc}[1]{{A_{#1}^{\rm NQC}}}
\newcommand{\omgmrg}[1]{{\omega_{#1}^{\rm mrg}}}
\newcommand{\apeak}[1]{{A_{#1}^{\rm peak}}}
\newcommand{\omgpeak}[1]{{\omega_{#1}^{\rm peak}}}

\newcommand{\avg}[1]{{\langle {#1} \rangle}}
\newcommand{\scri}{{\mathrsfs{I}}}
\newcommand{\stf}[1]{{\langle {#1} \rangle}}
\newcommand{\ord}{\mathcal{O}}
\newcommand{\f}{\frac}
\newcommand{\gsf}{\text{1GSF}}
\newcommand{\bk}{\text{0GSF}}
\newcommand{\el}{\ell}
\newcommand{\nn}{\nonumber}
\newcommand{\mbf}[1]{\mathbf{#1}}
\newcommand{\LOhat}{\!\hat{\,\mathbf{L}}_0}
\newcommand{\Lhat}{\!\hat{\,\mathbf{L}}_\text{N}}
\newcommand{\Jhat}{\hat{\mathbf{J}}_\text{N}}
\newcommand{\Sa}{\mbf{S}_1}
\newcommand{\Sb}{\mbf{S}_2}
\newcommand{\hs}{\hat{s}}
\newcommand{\hS}{\hat{S}}
\newcommand{\Lhatdot}{\dot{\hat{\,\mathbf{L}}}_\text{N}}
\newcommand{\LhatdotNew}{\dot{\hat{\,\mathbf{L}}}_{\text{N},\perp}}
\newcommand{\Sadot}{\dot{\mbf{S}}_1}
\newcommand{\Sbdot}{\dot{\mbf{S}}_2}
\newcommand{\Jdot}{\dot{\mbf{J}}}
\newcommand{\rdot}{\dot{\mbf{r}}}
\newcommand{\vdot}{\dot{\mbf{v}}}
\newcommand{\ud}{\mathrm{d}}
\newcommand{\Ldot}{\dot{\mbf{L}}_{\text N}}
\newcommand{\LN}{\mbf{L}_{\text N}}
\newcommand{\nnm}{\nonumber}
\newcommand{\mce}{\mathcal{E}}
\newcommand{\mcb}{\mathcal{B}}
\newcommand{\bayes}[2]{\mathcal{B}^{#1}_{#2}}

\newcommand{\dali}[0]{\texttt{TEOBResumS-Dalí}}
\newcommand{\gwecc}[0]{\texttt{gw\_eccentricity}}
\newcommand{\bilby}[0]{\texttt{bilby}}
\newcommand{\pseob}[0]{\texttt{pSEOBNRv5PHM}}
\newcommand{\dcssim}[1]{\texttt{dCS_0p#1}}
\def\TEOBResumS{\texttt{TEOBResumS}}
\def\TEOBResumSGIOTTO{\texttt{TEOBResumS-GIOTTO}}
\def\bajes{\texttt{bajes}}
\def\Boldtheta{\boldsymbol{\theta}}
\def\Boldd{\textbf{d}}
\def\TEOBResumSDali{\texttt{TEOBResumS-Dalí}}
\def\TEOBB{\texttt{pTEOBResumS}}
\def\cthree{c_{\rm{N}^3\rm{LO}}}
\def\mbhf{M_{\rm BH}^f}
\def\abhf{a_{\rm BH}^f}
\def\alphalm0{\alpha_{\ell m 0}}
\def\taulm0{\tau_{\ell m 0}}
\def\omegalm0{\omega_{\ell m 0}}
\def\lm{{\ell m}}

\def\Nevents{11}

\definecolor{cyan}{rgb}{0,0.9,0.9}
\definecolor{orange}{rgb}{0.9,0.5,0}
\definecolor{magenta}{rgb}{1,0,1}
\definecolor{purple}{rgb}{0.8,0.4,0.8}
\definecolor{gray}{rgb}{0.8242,0.8242,0.8242}
\definecolor{dodgerblue}{rgb}{0.12, 0.56, 1.0}

%% file: acronyms.tex
\newacro{adm}[ADM]{Arnowitt-Deser-Misner}
\newacro{bbh}[BBH]{binary black hole}
\newacroplural{bbh}[BBHs]{binary black holes}
\newacro{bh}[BH]{black hole}
\newacroplural{bh}[BHs]{black holes}
\newacro{bhns}[BHNS]{black hole-neutron star}
\newacro{bhpt}[BHPT]{black hole perturbation theory}
\newacro{bns}[BNS]{binary neutron star}
\newacro{bf}[BF]{Bayes factor}
\newacro{cbc}[CBC]{compact binary coalescence}
\newacro{ce}[CE]{Cosmic Explorer}
\newacro{da}[DA]{data analysis}
\newacro{dgi}[DGI]{Damour, Gopakumar and Iyer}
\newacro{et}[ET]{Einstein Telescope}
\newacro{eob}[EOB]{Effective-One-Body}
\newacro{eom}[EOM]{equations of motion}
\newacro{fd}[FD]{frequency domain}
\newacro{fft}[FFT]{Fast Fourier transform}
\newacro{ft}[FT]{Fourier transform}
\newacro{gw}[GW]{gravitational wave}
\newacroplural{gw}[GWs]{gravitational waves}
\newacro{gr}[GR]{general relativity}
\newacro{grb}[GRB]{gamma-ray burst}
\newacro{grhd}[GRHD]{general-relativistic hydrodynamics}
\newacro{gwosc}[GWOSC]{Gravitational Wave Open Science Center}
\newacro{gwtc1}[GWTC-1]{the first gravitational-wave transients catalog}
\newacro{gsf}[GSF]{Gravitational Self Force}
\newacro{hm}[HM]{Higher mode}
\newacroplural{hm}[HMs]{Higher modes}
\newacro{ifo}[IFO]{interferometer}
\newacro{imr}[IMR]{inspiral-merger-ringdown}
\newacro{im}[IMR]{inspiral-to-merger}
\newacro{kagra}[KAGRA]{Kamioka Gravitational Wave Detector}
\newacro{ligo}[LIGO]{Laser Interferometer Gravitational-Wave Observatory}
\newacro{lisa}[LISA]{Laser Interferometer Space Antenna}
\newacro{lr}[LR]{Light Ring}
\newacro{lso}[LSO]{Last Stable Orbit}
\newacro{lvc}[LVC]{LIGO-Virgo Collaboration}
\newacro{lvk}[LVK]{LIGO-Virgo-KAGRA}
\newacro{lo}[LO]{leading order}
\newacro{ns}[NS]{neutron star}
\newacroplural{ns}[NSs]{neutron stars}
\newacro{nr}[NR]{numerical relativity}
\newacro{nqc}[NQC]{next-to-quasicircular}
\newacro{nit}[NIT]{near-identity transformation}
\newacroplural{nit}[NITs]{near-identity transformations}
\newacro{nlo}[NLO]{next-to-leading order}
\newacro{nnlo}[NNLO]{next-to-next-to-leading order}
\newacro{n3lo}[N3LO]{next-to-next-to-next-to-leading order}
\newacro{n4lo}[N3LO]{next-to-next-to-next-to-next-to-leading order}
\newacro{ode}[ODE]{Ordinary Differential Equation}
\newacroplural{ode}[ODEs]{Ordinary Differential Equations}
\newacro{pn}[PN]{post-Newtonian}
\newacro{pm}[PM]{post-Minkowskian}
\newacro{pe}[PE]{parameter estimation}
\newacro{psd}[PSD]{power spectral density}
\newacroplural{psd}[PSD]{power spectral densities}
\newacro{pa}[PA]{post-adiabatic}
\newacro{qnm}[QNM]{quasi-normal mode}
\newacro{qc}[QC]{quasi-circular}
\newacro{snr}[SNR]{signal-to-noise ratio}
\newacro{spa}[SPA]{stationary-phase approximation}
\newacro{sxs}[SXS]{Simulating eXtreme Spacetimes}
\newacro{td}[TD]{time domain}
\newacro{ng}[NG]{Nect Generation}
\newacro{dcs}[dCS]{dynamical Chern-Simons}
\newacro{bgr}[beyond-GR]{beyond-\ac{gr}}
\newacro{spec}[SpEC]{Spectral Einstein Code}
\newacro{edgb}[EdGB]{Einstein-dilaton-Gauss-Bonnet}
\newacro{emri}[EMRI]{extreme mass-ratio inspiral}
\newacro{imri}[IMRI]{intermediate mass-ratio inspiral}
\newacro{gwtc}[GWTC]{Gravitational-Wave Transient Catalog}
\newacro{far}[FAR]{false alarm rate}
\newacro{ci}[CI]{credible interval}
\newacro{bs}[BS]{boson star}

%% file: pa_vs_ode_numbers.tex
\newcommand{\PAvsODETimeODEImri}{4.39}
\newcommand{\PAvsODETimePAImri}{1.29}
\newcommand{\PAvsODESpeedupImri}{3.4}
\newcommand{\PAvsODEDrMergerImri}{2.3 \times 10^{-3}}
\newcommand{\PAvsODEDphiMergerImri}{2.5 \times 10^{-2}}
\newcommand{\PAvsODEDrHandoffImri}{3.9 \times 10^{-5}}
\newcommand{\PAvsODEDphiHandoffImri}{4.7 \times 10^{-3}}
\newcommand{\PAvsODETimeODEBhnsLike}{0.55}
\newcommand{\PAvsODETimePABhnsLike}{0.24}
\newcommand{\PAvsODESpeedupBhnsLike}{2.3}
\newcommand{\PAvsODEDrMergerBhnsLike}{1.7 \times 10^{-4}}
\newcommand{\PAvsODEDphiMergerBhnsLike}{1.6 \times 10^{-3}}
\newcommand{\PAvsODEDrHandoffBhnsLike}{1.2 \times 10^{-6}}
\newcommand{\PAvsODEDphiHandoffBhnsLike}{1.2 \times 10^{-4}}
\newcommand{\PAvsODETimeODEBnsLike}{3.03}
\newcommand{\PAvsODETimePABnsLike}{0.50}
\newcommand{\PAvsODESpeedupBnsLike}{6.1}
\newcommand{\PAvsODEDrMergerBnsLike}{2.9 \times 10^{-3}}
\newcommand{\PAvsODEDphiMergerBnsLike}{1.6 \times 10^{-2}}
\newcommand{\PAvsODEDrHandoffBnsLike}{3.5 \times 10^{-6}}
\newcommand{\PAvsODEDphiHandoffBnsLike}{1.3 \times 10^{-3}}
\newcommand{\PAvsODETimeODEBbhPrec}{0.49}
\newcommand{\PAvsODETimePABbhPrec}{0.24}
\newcommand{\PAvsODESpeedupBbhPrec}{2.0}
\newcommand{\PAvsODEDrMergerBbhPrec}{2.3 \times 10^{-4}}
\newcommand{\PAvsODEDphiMergerBbhPrec}{1.5 \times 10^{-3}}
\newcommand{\PAvsODEDrHandoffBbhPrec}{5.6 \times 10^{-7}}
\newcommand{\PAvsODEDphiHandoffBbhPrec}{1.2 \times 10^{-4}}
\newcommand{\PAvsODESpeedupMin}{2.0}
\newcommand{\PAvsODESpeedupMax}{6.1}

%% file: v6cmp_numbers.tex
\newcommand{\VcmpCPU}{Intel Xeon W-2295}
\newcommand{\VcmpSystem}{Ubuntu 22.04, glibc 2.35}
\newcommand{\VcmpFactorVone}{0.63}
\newcommand{\VcmpFactorVfive}{0.79}
\newcommand{\VcmpVsVsixLo}{1.4}
\newcommand{\VcmpVsVsixHi}{5.2}
\newcommand{\VcmpVsVoneMin}{3.5}
\newcommand{\VcmpVsVoneMax}{26}
\newcommand{\VcmpVsVfiveMin}{1.7}
\newcommand{\VcmpVsVfiveMax}{7.3}
\newcommand{\VcmpOdeOverPaMin}{2.4}
\newcommand{\VcmpOdeOverPaMax}{3.7}

%% file: sweep_numbers.tex
\newcommand{\SwMmMedBbhLow}{2 \times 10^{-9}}
\newcommand{\SwDphiHandoffMaxBbh}{9 \times 10^{-4}}
\newcommand{\SwMmMedBnsLow}{2 \times 10^{-8}}
\newcommand{\SwDphiHandoffMaxBns}{2 \times 10^{-3}}
\newcommand{\SwMmMaxLowBound}{5 \times 10^{-7}}
\newcommand{\SwMmMedBbhSixSeven}{8 \times 10^{-8}}
\newcommand{\SwMmMedBbhAboveSeven}{8 \times 10^{-6}}
\newcommand{\SwMmMedBnsSixSeven}{8 \times 10^{-9}}
\newcommand{\SwMmMedBnsAboveSeven}{3 \times 10^{-7}}
\newcommand{\SwMmNBig}{4}
\newcommand{\SwMmBigEmin}{0.73}
\newcommand{\SwSpeedupBnsLow}{2.6}
\newcommand{\SwSpeedupBnsLowPfive}{1.8}
\newcommand{\SwSpeedupBnsLowPninetyfive}{3.5}
\newcommand{\SwSpeedupBnsLowEcc}{3.3}
\newcommand{\SwSpeedupBnsFiveSix}{1.7}
\newcommand{\SwSpeedupBnsAboveSeven}{0.8}
\newcommand{\SwTimeODEBns}{1.3}
\newcommand{\SwTimePABns}{0.50}
\newcommand{\SwFracBnsSubsecond}{99.7}
\newcommand{\SwFracBnsSubsecondLowEcc}{99.9}
\newcommand{\SwSpeedupBbhLow}{1.5}
\newcommand{\SwSpeedupBbhLight}{2.7}
\newcommand{\SwSpeedupBbhMaxLight}{7}
\newcommand{\SwMassBbhUnity}{23}
\newcommand{\SwSpeedupBbhHeaviest}{0.8}
\newcommand{\SwEccBbhUnity}{0.4}
\newcommand{\SwFracBbhSlower}{24}
\newcommand{\SwTimeBbhSlowerLo}{0.05}
\newcommand{\SwTimeBbhSlowerHi}{0.4}
\newcommand{\SwFitaa}{-19.4}
\newcommand{\SwFitab}{-3.49}
\newcommand{\SwFitac}{-6.35}
\newcommand{\SwFitad}{3.48}
\newcommand{\SwFitba}{16.5}
\newcommand{\SwFitbb}{1.73}
\newcommand{\SwFitbc}{-0.521}
\newcommand{\SwFitca}{-4.45}
\newcommand{\SwFitcb}{-0.0122}
\newcommand{\SwFitda}{0.396}
\newcommand{\SwFitNcycLo}{150}
\newcommand{\SwFitNcycHi}{7600}
\newcommand{\SwFitRightBbh}{95}
\newcommand{\SwFitRightBns}{99}
\newcommand{\SwPrecMmMaxBbhLow}{2 \times 10^{-5}}
\newcommand{\SwPrecMmMaxBbhHigh}{3 \times 10^{-3}}
\newcommand{\SwPrecMmMedBbhLow}{10^{-8}}
\newcommand{\SwPrecNBigBbh}{79}
\newcommand{\SwPrecNHighBbh}{552}
\newcommand{\SwPrecBigEminBbh}{0.63}
\newcommand{\SwPrecSpeedupBbhLow}{1.5}
\newcommand{\SwPrecSpeedupBbhHigh}{1.0}
\newcommand{\SwPrecMmMaxBnsLow}{6 \times 10^{-7}}
\newcommand{\SwPrecMmMaxBnsHigh}{10^{-3}}
\newcommand{\SwPrecMmMedBnsLow}{7 \times 10^{-8}}
\newcommand{\SwPrecNBigBns}{41}
\newcommand{\SwPrecNHighBns}{2241}
\newcommand{\SwPrecBigEminBns}{0.73}
\newcommand{\SwPrecSpeedupBnsLow}{2.3}
\newcommand{\SwPrecSpeedupBnsHigh}{2.1}

%% file: fd_numbers.tex
\newcommand{\FDmmBnsEcoHigh}{8.3 \times 10^{-8}}
\newcommand{\FDmmBbhEcoLow}{1.1 \times 10^{-6}}
\newcommand{\FDmmBbhEcoHigh}{1.4 \times 10^{-5}}
\newcommand{\FDmmBnsEcoLow}{8.3 \times 10^{-8}}
\newcommand{\FDresMrgMax}{2.8 \times 10^{-5}}